\def\expec#1{\langle#1\rangle}

\def\diag#1{\>{\rm diag}\{#1\}}
\def\nothing{\noindent\centerline{\,}}

\def\etal{{\frenchspacing\it et al.}}
\def\ie{{\frenchspacing\it i.e.}}
\def\eg{{\frenchspacing\it e.g.}}
\def\etc{{\frenchspacing\it etc.}}
\def\rms{{\frenchspacing r.m.s.}}

\def\rf#1;#2;#3;#4;#5 {\par#1, #3 {\bf #4}, #5 (#2). \par}

\def\beq#1{\begin{equation}\label{#1}}
\def\eeq{\end{equation}}
\def\beqa#1{\begin{eqnarray}\label{#1}}
\def\eeqa{\end{eqnarray}}
\def\eq#1{Eq.~(\ref{#1})}
\def\Eq#1{Eq.~(\ref{#1})}
\def\eqno#1{~(\ref{#1})}

\def\fig#1{Fig.~\ref{#1}}
\def\Fig#1{Fig.~\ref{#1}}

\def\A{{\bf A}}
\def\At{{\tilde{\bf A}}}
\def\B{{\bf B}}
\def\C{{\bf C}}
\def\D{{\bf D}}

\def\F{{\bf F}}

\def\I{{\bf I}}

\def\LL{{\bf\Lambda}}
\def\M{{\bf M}}
\def\N{{\bf N}}
\def\Nt{{\tilde{\bf N}}}
\def\NN{{\bf\Sigma}}
\def\P{{\bf P}}

\def\R{{\bf R}}
\def\S{{\bf S}}

\def\W{{\bf W}}

\def\0{{\bf 0}}
\def\zero{{\bf 0}}
\def\a{{\bf a}}
\def\b{{\bf b}}

\def\n{{\bf n}}
\def\nt{{\tilde{\bf n}}}
\def\r{{\bf r}}
\def\x{{\bf x}}
\def\y{{\bf y}}
\def\yt{{\tilde{\bf y}}}
\def\z{{\bf z}}
\def\xt{\tilde{\bf x}}
\def\yt{{\tilde{\bf y}}}

\def\rh{{\widehat{\bf r}}}
\def\err{{\bf\varepsilon}}
\def\l{\ell}

\def\tobs{t_{obs}}

\def\tr{\hbox{tr}\>}

\def\muK{{\rm \mu K}}
\def\Wh{\widehat{W}}

\def\simpropto{\,{\lower3.0pt\hbox{$\propto$}\atop\raise1.0pt\hbox{$\sim$}}\,}
\def\simlt{\lesssim}
\def\simgt{\gtrsim}

\def\tento#1{\times 10^{#1}}

\def\Cnoise{C^{noise}}
\def\leff{\l^{eff}}
\def\fsky{f_{sky}}
\def\sigmacmb{\sigma_{cmb}}

\documentstyle[prd,twocolumn,aps,epsf,rotate]{revtex}
\begin{document}




\preprint{IASSNS-AST 97/666}

\title{CMB MAPPING EXPERIMENTS: A DESIGNER'S GUIDE}

\author{Max Tegmark\thanks{Hubble Fellow.}}

\address{Institute for Advanced Study, Princeton, NJ 08540; max@ias.edu}


\date{Submitted May 24, 1997; accepted July 29, 1997}

\maketitle

\begin{abstract}
\noindent{\bf Abstract:} 
We apply state-of-the art data analysis methods to a number of 
fictitious CMB mapping experiments, including $1/f$ noise,
distilling the cosmological information 
from time-ordered data to maps to power spectrum estimates, 
and find that in all cases, the resulting error bars can we well 
approximated by simple and intuitive analytic expressions.
Using these approximations, we discuss how to maximize the scientific
return of CMB mapping experiments given the practical constraints at hand,
and our main conclusions are as follows.
(1) For a given resolution and sensitivity,
it is best to cover a sky area such that the signal-to-noise ratio
per resolution element (pixel) is of order unity.
(2) It is best to avoid excessively skinny
observing regions, narrower than a few degrees.
(3) The minimum-variance mapmaking method can reduce the effects of 
$1/f$ noise by a substantial factor, but only if the scan pattern
is thoroughly interconnected. 
(4) $1/f$ noise produces a $1/\l$ contribution to the angular power 
spectrum for well connected single-beam scanning, as compared to virtually white 
noise for a two-beam scan pattern such as that of the $MAP$ satellite.

\end{abstract}


\makeatletter
\global\@specialpagefalse
\def\@oddfoot{
\ifnum\c@page>1
  \reset@font\rm\hfill \thepage\hfill
\fi
\ifnum\c@page=1
  {\sl 
Available in color from 
h t t p://www.sns.ias.edu/$\tilde{~}$max/strategy.html} \hfill\\
\fi
} \let\@evenfoot\@oddfoot
\makeatother



\section{INTRODUCTION}

Over the next decade, precision measurements of the
cosmic microwave background (CMB) are likely to 
radically tighten existing constraints on cosmological models.
Although some upcoming experiments, {\eg}, the 
NASA MAP Satellite, already have their design and observing
strategy essentially frozen in, many others do not, and face
important tradeoffs between figures of merit such as 
resolution, sky coverage, frequency coverage and sensitivity. 
For instance, is it better to concentrate a given amount of 
observing time on a small patch, thereby improving the
signal-to-noise per pixel, or to map a large area with lower accuracy?
The purpose of this paper is to investigate how such
tradeoffs affect the accuracy with which cosmological models can
be constrained, thereby providing some guidance for
observers attempting to maximize the scientific ``bang for the buck"
of their experiments.

\subsection{From maps to cosmology}

The approach taken with the first CMB experiments was to use 
numerical likelihood or Monte Carlo  calculations to assess the
accuracy with which various parameters could be measured from the 
data. It has gradually become clear that although such calculations 
are useful {\it post hoc}, to compute accurate error bars 
once the experiment has taken place and the data set is in hand,  
simple and intuitive analytic approximations exist that
are often accurate enough for studying the effects of design tradeoffs.
For instance, it was shown that the effect of incomplete sky coverage
is well approximated by two simple effects:
to increase the sample variance by a factor $1/\fsky$
\cite{Scott94}, where $\fsky$ is the fraction of the sky area that 
is observed, and to smear out features in the power spectrum
on a scale $\Delta\l\sim 1/\Delta\theta$
\cite{window}, where $\Delta\theta$ is the size of the patch
(in radians) in its narrowest direction.
In a similar spirit, Knox showed that 
the effect of uniform instrumental noise could be accurately modeled as
simply an additional random field on the sky, with an angular power
spectrum given by \cite{Knox95,wiener}
\beq{ClNoiseEq}
\Cnoise_\l 
= {\Omega\sigma^2\over N B_\l^2}
= {\Omega s^2\over\tobs B_\l^2}
= {\fsky \over w B_\l^2},
\eeq
and we give a detailed proof of this in Appendix A.
Here $\sigma$ is the {\rms} noise in each of the $N$ pixels,
the solid angle covered is $\Omega=4\pi\fsky$, $s$ is the 
detector sensitivity in units $\muK\,s^{1/2}$, $\tobs$ is the total 
observation time, and $w$ is the raw sensitivity measure
defined by \cite{Knox95} 
\beq{wDefEq}
w^{-1}\equiv {4\pi\sigma^2\over N} = {4\pi s^2\over\tobs}.
\eeq
$B_\l$ is the experimental beam function, which for a Gaussian 
beam with standard deviation $\theta_b$\footnote{
The FWHM (full-width-half-maximum) is given by
FWHM=$\sqrt{8\ln 2}\theta_b$.}
is well approximated by
\beq{BlEq}
B_\l = e^{-\theta_b^2\l(\l+1)/2}.
\eeq
Thus early estimates 
of how accurately cosmological parameters
could be measured based on Monte Carlo  maps 
(\eg \cite{Hinshaw95})
could be substantially accelerated.
A further simplification was achieved by altogether
eliminating the likelihood minimization (performed 
say by simulated annealing \cite{Knox95}), and computing 
the attainable error bars directly from the power spectrum and its
derivatives \cite{JKKS1}. This procedure involves the 
formalism of the Fisher Information Matrix 
(described in detail in \cite{karhunen}), and has now been used to 
study the accuracy with which about a dozen cosmological 
parameters can be simultaneously 
measured by {\it MAP} and the ESA 
{\it Planck} mission \cite{JKKS2,parameters,Zaldarriaga97},
going substantially beyond the 
obvious conclusions that it helps to
increase the sky and frequency coverage, the resolution and the
sensitivity. It was found that by increasing the angular resolution
to FWHM$\ll 1^{\circ}$, thereby measuring the power spectrum 
well beyond the first ``Doppler peak",
much of the degeneracy between different parameters that
had been termed ``cosmic confusion'' \cite{Confusion}
could be lifted, with 
{\it Planck} measuring most parameters to within a few percent.
Measuring polarization as well was found to improve the accuracy by 
a further factor of two assuming that foreground and systematics
problems could be controlled \cite{Zaldarriaga97}.
Using the same method, a number of experimental design issues for 
both single-dish experiments
and interferometers have been discussed with the attention limited
to measuring the density parameter $\Omega$ \cite{MH97}
(``weighing the Universe'') and 
the observability of a second Doppler peak \cite{Hobson97}.

\subsection{From time-ordered data to maps}

All the above-mentioned results focused on the link between 
completed CMB maps and cosmological constraints.
In the presence of detector $1/f$ noise,
however, it is important to pay attention also to the previous
step in the data-analysis pipeline, where the time-ordered data (TOD)
is reduced to a map. Handy approximations for the impact of 
$1/f$ noise when circular scans are averaged have been 
derived \cite{Janssen96}, and it is clear that 
the scan strategy (by which we mean not merely how many times
different pixels are observed, but also in what order) has a substantial 
impact on the attainable noise levels in the map. 
It has been argued \cite{Wright96} that it is desirable to have a
scan strategy that is as ``connected" as possible, where each pixel
is scanned through in many different directions. 

\subsection{New data-analysis techniques}

Substantial progress has recently been made on the issue
of how to analyze a given data set.
Computationally feasible 
methods are now available for reducing 
data sets as large as those of the upcoming satellite missions
from time-ordered data to maps and 
from maps to power spectra and cosmological parameter constraints
in a way that destroys no 
cosmological information, in the sense that  
parameters can be measured just as accurately
as they could with a (computationally unfeasible)
brute force likelihood analysis of the entire time ordered
data set. 
A recent clever implementation of the minimum-variance 
method for reducing TOD to maps \cite{Wright96} is
both feasible and lossless in this sense \cite{mapmaking}
(all the cosmological information from the TOD is distilled
into the map with nothing leaking out of the pipeline).
A feasible and lossless power spectrum estimation has also 
been found \cite{cl,Knox97} for the case of Gaussian fluctuations.
The signal-to-noise eigenmode method 
(see \cite{Bond95,Bunn+Sugiyama,Bunn95,karhunen,Jaffe97} and 
references therein) offers a feasible 
and lossless way of constraining parameters directly 
from maps as long as the number of pixels $n\simlt 10^4$,
as do the orthogonalized spherical harmonic 
\cite{Gorski94} and brute-force \cite{brute,Hinshaw96} methods.

\subsection{Outline}

In this paper, we will adopt an approach to experimental design
which combines the accuracy of these new numerical methods
with the intuitive understanding of the
analytical approximations. 
This has essentially not been done before. For instance, the 
published $1/f$ approximations \cite{Janssen96} were not based on the 
lossless mapmaking method \cite{Wright96,mapmaking}, but on a straight pixel 
averaging which can be improved upon in many situations, and 
the resulting angular power spectrum of the noise
was not computed exactly given $1/f$ noise, 
merely estimated with Monte Carlo simulations \cite{Janssen96}.
Similarly, the above-mentioned 
sample variance approximation was derived 
assuming a Gaussian autocorrelation function \cite{Scott94}, although
as we will see, it is readily generalized to a signal-to-noise or
power spectrum analysis.
We will present a number of
worked examples, using the above-mentioned lossless data analysis methods, 
and show how in each case, these results can be accurately
matched by simple approximations. 
We then use these approximations to arrive at rules of thumb for
experimental design.
Section 2 discusses the effect of varying 
four attributes of a map; its size, shape, sensitivity and resolution.
(For a discussion on the best choice of frequency channels with
regard to foreground removal, see {\eg} 
\cite{Brandt94,Dodelson97,wiener}.)
Section 3 discusses the preceding mapmaking step, and 
how two attributes of the scan pattern 
(the $1/f$ noise level and the amount of interconnectedness in the
scan pattern) affect the noise power spectrum in the map.


\section{FROM MAP TO COSMOLOGY}

In this section, we analyze a number of different types of maps
with the signal-to-noise eigenmode method and the lossless power
spectrum method, focusing on the effect of varying 
the map size, shape, sensitivity and resolution.

\subsection{Signal-to-noise eigenmodes: demystifying the black box}

The signal-to-noise (S/N) eigenmode method 
distills the information content of a CMB map into a set of
mutually exclusive and collectively exhaustive chunks which
have a number of properties that make them useful for measuring 
the CMB power spectrum and constraining cosmological models.
Although it has traditionally been a ``black box'' method, 
where all the details are hidden in the numerical diagonalization of a 
large matrix, we will see below that the workings of this box
are in fact easy to understand both qualitatively and quantitatively
by making some simple approximations.\footnote{An interesting step in this
direction was a handy approximation
for the special case of the MSAM experiment chopping\cite{Knox96}.}

\subsection{A minimalistic review of the S/N method}

The S/N eigenmode method was introduced into
CMB data analysis by Bond \cite{Bond95} and Bunn \cite{Bunn95},
who both reinvented the method independently. 
It is a special case of the Karhunen-Lo\`eve 
method \cite{Karhunen47}, and since our focus here is not on 
data analysis methods but on experimental design, our review below is 
very brief and the interested reader is referred to other recent papers
\cite{karhunen,Jaffe97} for method details.
Suppose the CMB map is pixelized into $N$ pixels whose center positions in
the sky are given by the unit vectors 
$\rh_1,\rh_2,...,\rh_N$. 
The map consists of $N$ numbers ${\tilde x}_i\equiv x_i+n_i$, where
$x_i\equiv \delta T(\rh_i)$ are the true sky temperatures and $n_i$ are the instrumental 
noise contributions. We group these numbers into $N$-dimensional vectors
$\xt$, $\x$ and $\n$, respectively,  so $\xt=\x+\n$.
The signal $\x$ and noise $\n$ are assumed to have zero mean
($\expec{\x}=\expec{\n}=\zero$), to be uncorrelated
($\expec{\x\n^t}=\zero$), and to have a multivariate Gaussian probability
distribution with covariance matrices $\S\equiv\expec{\x\x^t}$
and $\N\equiv\expec{\n\n^t}$. The data covariance matrix is thus
$\C\equiv\expec{\xt\xt^t}=\S+\N$.
The signal-to-noise eigenmodes are the $N$ vectors $\b_i$
satisfying the generalized eigenvalue equation
\beq{SNeq}
\S\b_i = \lambda_i\N\b_i.
\eeq
Grouping them together as the columns of the $N\times N$
matrix $\B$, one computes a new data vector $\y\equiv\B^t\xt$.
These $N$ numbers $y_i$ are the above-mentioned 
information chunks. They are mutually exclusive in
the sense that they are uncorrelated 
($\expec{y_i y_j}=\b_i^t\C\b_j = [1+\lambda_i]\delta_{ij}$)
and collectively exhaustive in the sense that they
retain all the information from the original data set
(since $\xt$ can be recovered by computing $\B^{-t}\y$, as
$\B$ is invertible).
Moreover, the eigenvalues $\lambda_i$
can be interpreted as signal-to-noise ratios 
for the coefficients $y_i$, 
and sorting them by decreasing signal-to-noise, $y_1$ 
can be shown to contain the most information about the 
power spectrum normalization, followed by $y_2$, $y_3$, {\etc}
Typically, the bulk of the coefficients are so noisy that they
can be thrown away without appreciable loss of cosmological information, 
and such data compression has the advantage of greatly accelerating
subsequent analysis such as likelihood computations, where the CPU time
typically scales as the cube of the size of the data set.

The expectation value $\expec{y_i^2}$ generally 
equals a noise term stemming from $\N$ plus a linear combination
of $\delta T_\l^2 \equiv \l(\l+1)C_\l/2\pi$, 
where the weights given to the different $\delta T_\l^2$ are
denoted $W^{(i)}_\l$, the {\it window function}.
Here $C_\l$ is the customary angular power spectrum,
and the window functions are given by (\eg \cite{window})
\beq{WindowEq}
W^{(i)}_\l \propto {2\l+1\over\l(\l+1)}B_\l^2\sum_{j=1}^N\sum_{k=1}^N
\B_{ji}\B_{ki} P_\l(\rh_j\cdot\rh_k),
\eeq
where $P_\l$ denotes Legendre polynomials. 
$W$ is normalized so that $\sum_\l W^{(i)}_\l=1$, so we can think 
of $y_i^2$ as measuring a weighted average of the power
spectrum coefficients $\delta T_\l^2$, with the window function giving
the weights. 
As the examples below will illustrate, the coefficients 
$\y_i$ generally have the additional advantage of being fairly localized
in the Fourier (multipole) domain, by which is meant that they
have narrow window functions, and this makes them useful for 
band power measurements.

\subsection{Case study 1: round maps}

Let us first consider a CMB map with an angular resolution $\theta_b$
covering the sky area
within an angle $\theta$ from some given point. 
For $\theta\ll 1$, this region will simply be a (rather flat) disk
of radius $\theta$, whereas $\theta=\pi$ gives a full sky map.
The sky fraction covered is 
\beq{junkEq1}
\fsky=\sin^2{\theta\over 2}.
\eeq
We discretize the map into $N$ equal-area pixels which we
assume to have uncorrelated Gaussian noise with an {\rms}
amplitude $\sigma$. To keep things simple, we use a flat 
fiducial power spectrum $C_\l\propto 1/\l(\l+1)$
with a $Q=30\muK$ quadrupole normalization, corresponding to
$\delta T_\l=(12/5)^{1/2}Q\approx 47\muK$, a ball park figure
for recent degree-scale measurements.

\begin{figure}[phbt]
\centerline{\rotate[r]{\vbox{\epsfxsize=18cm\epsfbox{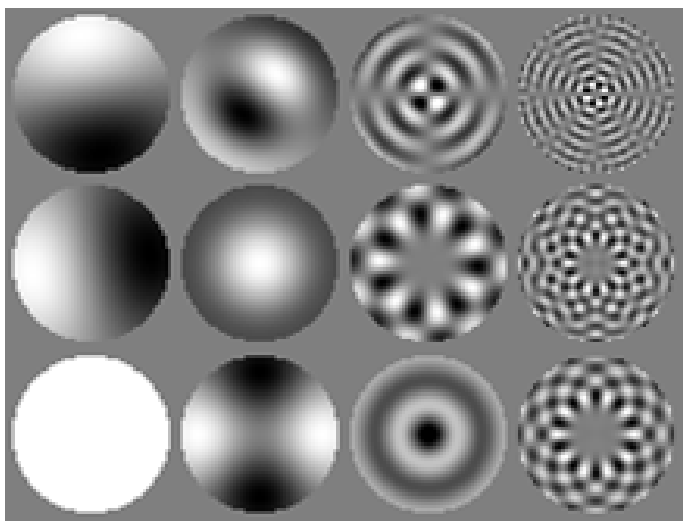}}}}
\caption{
The signal-to-noise eigenmodes are plotted
for a circular sky patch of $10^\circ$ diameter 
using a flat fiducial power spectrum.
The modes plotted are, from left to right, top to bottom, 
1, 2, 3, 4, 6, 10, 30, 50, 100, 
150, 300 and 500.
}
\label{DiskModesFig}
\end{figure}

\begin{figure}[phbt]
\centerline{{\vbox{\epsfxsize=16cm\epsfbox{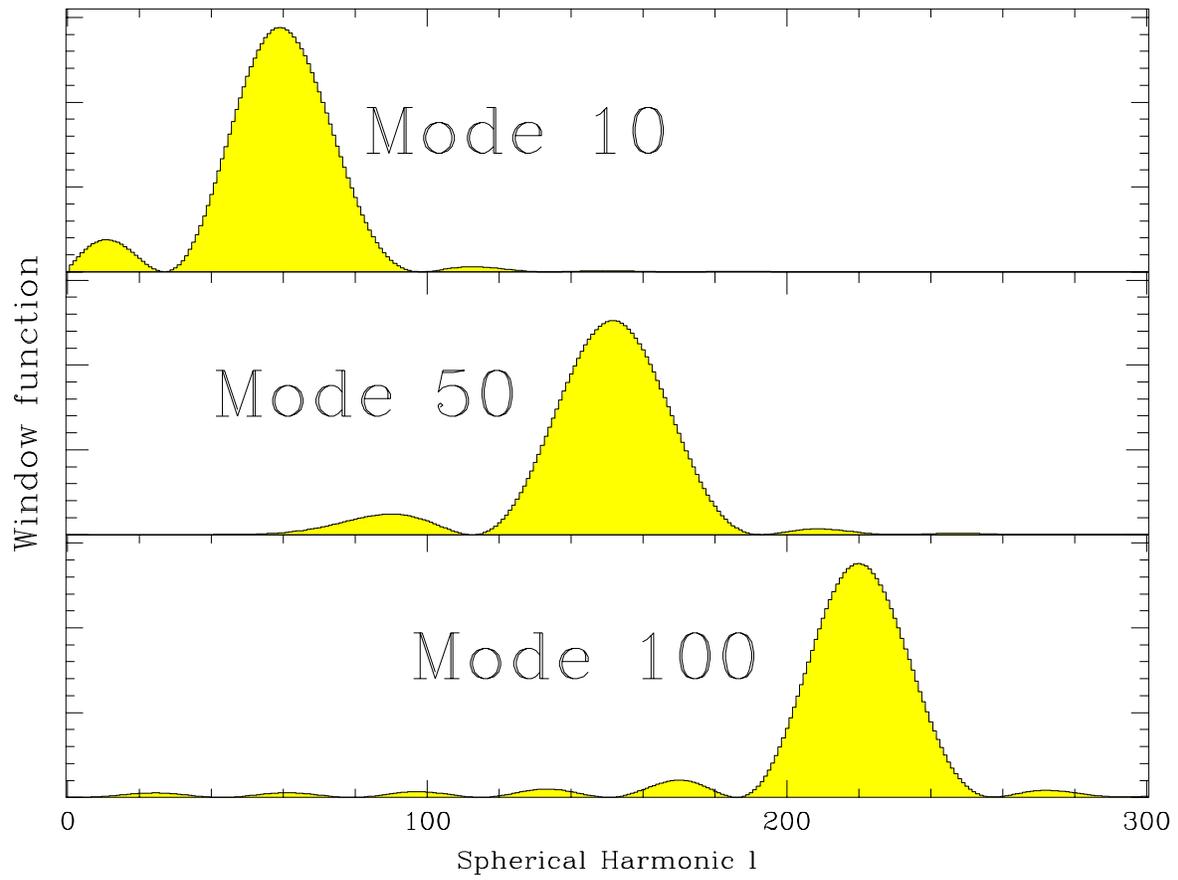}}}}
\caption{
The window functions are shown for three of the signal-to-noise
eigenmodes from the previous figure.
}
\label{LocationFig}
\end{figure}

\Fig{DiskModesFig} shows the eigenmodes for the case 
$\theta=5^\circ$. Some of the corresponding window functions are 
plotted in \fig{LocationFig}, and the eigenvalues $\lambda_i$ are shown 
in \fig{snfitsFig}. As we will now describe, the contents of all
of these figures could have been approximately predicted without ever
carrying out the full numerical calculation.

\subsubsection{Eigenmodes:}

Let us first consider the eigenmodes.
The exact choice of pixelization is clearly irrelevant as long as 
the pixel separation is much smaller than the beamwidth, so let us simplify the
problem by considering an infinitely fine pixelization, where the 
eigenmodes are smooth functions $b_i(\rh)$.
The spherical harmonics $Y_{\l m}$ are eigenfunctions of the Laplacian $\Delta$ 
with eigenvalues $\l(\l+1)$, so multiplying by $\l(\l+1)$
in the Fourier (multipole) domain is equivalent to applying the
angular Laplace operator on the sphere. 
When the fiducial power spectrum is 
$C_\l\propto  1/\l(\l+1)$, we can thus think of 
the signal covariance matrix in \eq{SNeq}
as essentially
$\S\propto\Delta^{-1}$.
Since $\N\propto\I$, the eigenmodes are thus basically 
eigenfunctions of the Laplacian.
When $\theta\ll 1$, sky curvature is negligible and 
this reduces to the 2D Helmholtz equation. 
For the circular case at hand, the solutions are well-known to
correspond to Bessel functions:
\beq{BesselEq}
b_{\l m}(\r) \propto J_m(k_\l r) e^{i m\phi},
\eeq
where $\r=(x,y)=r(\cos\phi,\sin\phi)$.
Thus each mode is specified by some 
integer $m$ and some radial wave number $k_\l$.
This is verified by our numerical results.
Since the discrete eigenvectors of \eq{SNeq} are orthogonal
when $\N\propto\I$, the combination of 
$m$ and $k_\l$ will be such that
the functions $b_{\l m}(\r)$ are orthogonal as well. 

\subsubsection{Window functions:}
   
\Fig{DiskModesFig} also shows that the larger mode numbers
tend to oscillate more. This reflects the fact that the 
window functions probe increasingly small scales (large $\l$)
as the mode number increases, which is 
more clearly illustrated in \fig{LocationFig}.
This is quite a general property of the CMB S/N method
\cite{Bunn95}, and holds because
whereas the signal $C_\l$ generally decreases with $\l$, the
noise power $\Cnoise_\l$ stays constant and eventually 
increases.\footnote{
In Section~\ref{NoisePowerSec}, we will see that $1/f$ noise can in
fact produce a falling $\Cnoise_\l$. Hovever, it generally falls no faster than
$\l^{-1}$ whereas the CMB signal falls as $\l^{-2}$, so our conclusion 
remains unaffected.
}
Thus the mode with the highest S/N-ratio will probe the 
largest scales to which the map is sensitive, the runner up
will be the largest scale mode remaining (which is uncorrelated with
the first one), {\etc}

The finite size of the survey places a crude lower limit on the 
width of the any window function \cite{window}: 
\beq{UncertaintyEq}
\Delta\l\simgt 1/\Delta\theta,
\eeq
where $\Delta\theta$ is the angular extent of the survey in the smallest direction, 
in our case $\Delta\theta\sim 2\theta$
(a more careful discussion of this is given in Section~\ref{WindowSec}).
This limit is typically attained with the decorrelated quadratic method
\cite{cl}, whereas the S/N-method sometimes does significantly worse.
We find that the mean of a window function, which we will denote $\expec{\l}$
or $\leff$ and define by
\beq{leffDefEq}
\leff_k\equiv\sum_\l \l\,W_\l^{(k)}
\eeq
is numerically well approximated by 
\beq{leffApproxEq}
\leff_k \approx \sqrt{k\over\fsky}.
\eeq
This can be understood as follows.
If we Fourier transform a finite patch of a homogeneous
random field, 
the Fourier coefficients become 
correlated over a {\it coherence volume} in Fourier space 
whose size is roughly the inverse of that of the patch --- this is 
a well-known effect in the context of galaxy surveys \cite{FKP}.
The situation is quite analogous with spherical harmonics \cite{Scott94,Hobson97}:
the number of multipole coefficients that become correlated 
are roughly $1/\fsky$. 
There are $\l_{max}^2$ multipoles $Y_{\l m}$ with $\l<\l_{max}$,
so one expects to be able to form roughly $\fsky\l_{max}^2$ 
uncorrelated linear combinations of them. 
Since the S/N-coefficients are all by construction uncorrelated, 
one therefore expects there to be of order 
$k=\fsky\l^2$ of them probing scales out to $\leff\sim\l$, in
agreement with \eq{leffApproxEq}.

\subsubsection{The signal-to-noise eigenvalues}

\begin{figure}[phbt]
\centerline{{\vbox{\epsfxsize=16cm\epsfbox{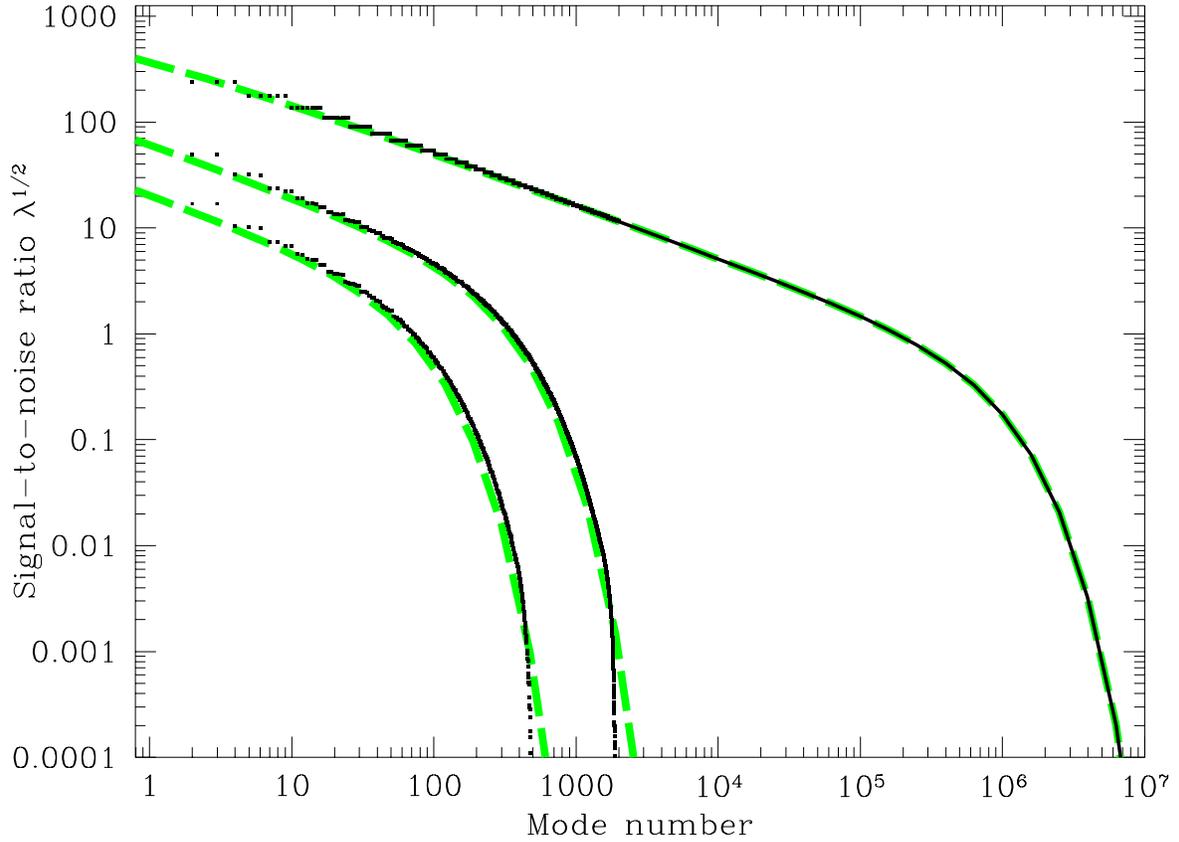}}}}
\caption{
Understanding signal-to-noise eigenmodes: 
exact calculations and approximations.
The signal-to-noise ratio $\lambda^{1/2}$ is plotted (dots)
for three exact numerical calculations together with 
the approximation of \eq{LambdaApproxEq} (dashed lines).
From top to bottom, the three cases correspond to 
complete sky coverage with $w^{-1}=7\tento{-15}$,
a disk of diameter $10^\circ$ with $w^{-1}=9\tento{-16}$,
and a $5^\circ$ disk with $w^{-1}=2\tento{-15}$.
}
\label{snfitsFig}
\end{figure}

As mentioned, a S/N-coefficient measures a weighted average 
of the power spectrum. As long as $\Delta\l\ll\l$ and the power
spectrum lacks sharp 
features, this average is well approximated by the power
at $\leff$, and we arrive at the useful approximation
\beq{LambdaApproxEq}
\lambda_k\approx\left.{C_\l\over \Cnoise_\l}\right|_{\l=\leff_k},
\eeq
where $\leff_k$ and $\Cnoise_\l$ are given by 
equations\eqno{leffApproxEq} and\eqno{ClNoiseEq}, respectively. 
As shown in \fig{snfitsFig}, this approximation is generally 
quite accurate. Symmetries tend to cause 
groups of modes to be degenerate, with identical eigenvalues, causing 
horizontal lines to be visible for the first modes. 
For the all-sky case, the $2\l+1$ multipoles corresponding to different
$m$-values are degenerate, and for azimuthally symmetric regions, the eigenvalues
come in pairs corresponding to a sine and a cosine mode.
At the opposite end, the very last modes are seen to contain
even less signal than predicted by \eq{LambdaApproxEq}. 
This is because the effect of discrete pixelization becomes noticeable
when the number of modes approaches the number of pixels.

We have tested our approximation for maps of a garden variety of
shapes and sizes, and in all cases find an accuracy comparable to
that in \fig{snfitsFig}.
Because it is both accurate and computationally trivial, 
it is a useful alternative to full-blown simulations 
and S/N-calculations
when studying experimental design issues.

\subsection{Lesson 1: how to choose the map size}
\label{SizeSec}

Above we found that the accuracy with which band powers could be
measured was accurately fit by the simple approximation
given by equations\eqno{ClNoiseEq} and\eqno{leffApproxEq}.  
Let us now use this result to address the following question:
given a fixed amount of observing time, how large a sky area should 
one spread it over? It is better to scan as large an area as practically 
feasible, or to map a smaller patch with a lower noise per 
pixel (resolution element)?

\subsubsection{How accurately can you measure band power?}

Let $\bar{C}_\l$ denote the power $C_\l$ averaged
over a multipole band $\l-L/2\le\l\le\l+L/2$, {\ie}, a band of width 
$L$ centered on $\l$.
How accurately can we measure the band power $\bar{C}_\l$?
\Eq{LambdaApproxEq} showed that as long as $L\gg\Delta\l$, 
each S/N eigenmode whose window function fell into this band 
would measure the power with a signal-to-noise ratio
$\lambda\approx C_\l/\Cnoise_\l$, which corresponds to measuring 
the band power with an {\rms} error $\sqrt{2}(C_\l+\Cnoise_\l)$
since the S/N-coefficient has a Gaussian distribution. 
(These two terms correspond to sample variance and noise variance, 
respectively.)
From our mode counting above, we know that there 
are ${\cal N}\sim(2\l+1)L\fsky$ such eigenmodes probing
the multipole band, so since they are by construction uncorrelated,
the {\rms} error simply drops by a factor $\sqrt{\cal N}$ when we use all
of them, giving
\beq{DeltaClEq1}
\Delta \bar{C}_\l \approx\sqrt{2\over(2\l+1)L\fsky}\left[C_\l + \Cnoise_\l\right]
\eeq
This is to be compared with the equation 
\beq{DeltaClEq2}
\Delta C_\l \approx\sqrt{2\over(2\l+1)\fsky}\left[C_\l + \Cnoise_\l\right],
\eeq
which has frequently appeared in the literature and follows if we naively
set $L=1$ in \eq{DeltaClEq1}. This is of course not legitimate when $\fsky<1$,
since \eq{DeltaClEq1} is only valid when $L\gg\Delta\l$, which
is just another way of saying that one cannot measure 
an individual multipole $C_\l$ alone when faced with incomplete sky coverage.
There is, nonetheless, a sense in which \eq{DeltaClEq2} can 
be used, with the appropriate precautions: as long as the power spectrum is smooth
enough to be featureless on the scale $\Delta\l$, 
calculations assuming that we can make uncorrelated measurements
of the individual multipoles with standard deviation 
given by \eq{DeltaClEq2} will always give the right answer.
For example, with this assumption, 
\eq{DeltaClEq2} can be used to derive \eq{DeltaClEq1}. Also, the
Fisher information matrix $\F$, which determines the accuracy with which 
cosmological parameters $\theta_1,\theta_2,...$ can be measured
\cite{karhunen}, is correctly given by 
\beq{FisherEq}
\F_{ij}=\sum_\l(\Delta C_\l)^{-2}
{\partial C_\l\over\partial\theta_i}{\partial C_\l\over\partial\theta_j}
\eeq
in this case, where $\Delta C_\l$ is given by \eq{DeltaClEq2}.

\subsubsection{Maximizing the accuracy}

Let us now vary $\fsky$ to minimize the measurement error
on the band power $\bar{C}_\l$. Substituting 
\eq{ClNoiseEq} into \eq{DeltaClEq1} gives
\beq{DeltaClEq3}
\Delta {\bar C}_\l\propto \fsky^{-1/2}\left[C_\l + {\fsky\over w B_\l^2}\right].
\eeq
Requiring the derivative of this with respect to $\fsky$ to vanish shows that 
the best choice of $\fsky$ is 
\beq{fskyOptEq}
\fsky = w B_\l^2 C_\l  = {N B_\l^2 C_\l\over 4\pi\sigma^2}.
\eeq
Substituting this back into \eq{ClNoiseEq}, we obtain $\Cnoise_\l = C_\l$, 
so we see that this choice corresponds to making the 
noise and sample variance contributions equal.

This choice of $\fsky$ depends on the multipole $\l$ that we are trying to
measure, so which $\l$ should be tailor the experiment for?
We argue that the natural choice is 
$\l\sim\l_b\equiv 1/\theta_b$, the scale set by the beam resolution, 
for the following reasons:
\begin{enumerate}
\item If one focuses on $\l\ll\l_b$, using the narrow 
beam is like throwing pearls before swine, since one would obtain
about as good results even with inferior angular 
resolution.
\item If one focuses on $\l\gg\l_b$, the beam factor 
$B_\l^2$ will be exponentially small and the resulting error bars
will be exponentially large.
\end{enumerate}
To obtain a rule of thumb for choosing the map size, we will therefore 
maximize the sensitivity to the scale $\l\sim\l_b$, {\ie}, 
where the experiment has
its strongest comparative advantage over others. Since $B_{\l_b}\sim 1$, this
gives simply $\fsky\sim w C_{\l_b}$.
Let us translate this into a more intuitive expression.
In terms of a flat band power $Q^2$, the power spectrum at $\l=\l_b$ is
by definition
\beq{FlatClEq}
C_\l = {24\pi\over 5}{Q^2\over\l_b(\l_b+1)}\sim{24\pi Q^2\theta_b^2\over 5}.
\eeq
If we divide the map area $4\pi\fsky$ into $N$ pixels of area
$FWHM^2$, $FWHM\equiv\sqrt{8\ln 2}\theta_b$, then 
equations\eqno{wDefEq} and\eqno{FlatClEq} together with our 
result $\fsky\sim w C_{\l_b}$ gives
\beq{NapproxEq}
N = {4\pi\fsky\over FWHM^2} \sim 
{3\pi\over 5\ln 2}{Q^2\tobs\over s^2} \sim 3{Q^2\tobs\over s^2}
\eeq
In other words, {\it for a given resolution and sensitivity,
it is best to cover a sky area of order the beam area times 
the signal-to-noise factor $Q^2\tobs/s^2$}.
Again using \eq{wDefEq}, this tells us that the 
the noise per pixel should be of order
\beq{sigmaApproxEq}
\sigma\sim 2 Q.
\eeq
Since the {\rms} CMB fluctuation $\sigmacmb$ in each pixel,
given by
\beq{PixelSignalEq}
\sigmacmb = \left[\sum_\l {2\l+1\over 4\pi}B_\l^2 C_\l\right]^{1/2},
\eeq
differs from $Q$ only by a logarithmic factor of order a few for 
typical angular resolutions and cosmological models,
we arrive at the following useful rule of thumb:
\begin{itemize}
\item {\it Choose the map size such that the signal-to-noise ratio per pixel, 
$\sigmacmb/\sigma$, is of order unity}.
\end{itemize}
Thus if an experiment has a noise level
per resolution element (pixel) of $\sigma\ll 50-100\muK$, it will  
be dominated by sample variance, and better results can be obtained
by spreading the scan out over a larger sky patch.
(Since $\fsky$ cannot exceed unity, it of course still makes 
sense to aim for lower noise levels for full-sky experiments, such as 
for instance the upcoming satellite missions.)
This rule of thumb agrees well with detailed calculations performed 
for the specific case of the MSAM2 experiment \cite{Knox96}.

\subsection{Case study 2: oblong maps}

Above we used the fact that both noise and sample variance depends 
only on the {\it area} of a map to determine the best choice of map size.
Changing the shape while keeping the area fixed leaves the variance 
unchanged but affects the width of the
window functions, the spectral resolution. 
To study this in more detail, we will now study the effect of
elongating a map, returning to a discussion of how to best choose the
map shape in the next section. 
Consider a small rectangular map of size $\theta_x\times\theta_y$,
where $\theta_x\le\theta_y\ll 1$, so that we can neglect the effect of sky 
curvature. 

\subsubsection{The eigenmodes}

As discussed above, we expect the signal-to-noise eigenmodes
to be eigenfunctions of the Laplacian, which for rectangular symmetry take the form
\beq{RectEigenEq}
b_{mn}(x,y) \propto \cos(k_x x+\alpha)\cos(k_y y +\beta),
\eeq
where the wave numbers $k_x$ and $k_y$ and the phases $\alpha$ and $\beta$
are such that the modes are orthogonal.
(We are using coordinates where the map is centered on the north pole, 
so $\rh\approx (x,y,1)$, $|x|\le\theta_x$, $|y|\le\theta_y$.)
\Eq{RectEigenEq} is verified by our numerical computations, and illustrated in
\fig{shapeFig} and \fig{positionFig}, where six sample modes are
plotted together with their window functions.

\subsubsection{The window functions}

\label{WindowSec}

\begin{figure}[phbt]
{{\vbox{\epsfxsize=16cm\epsfbox{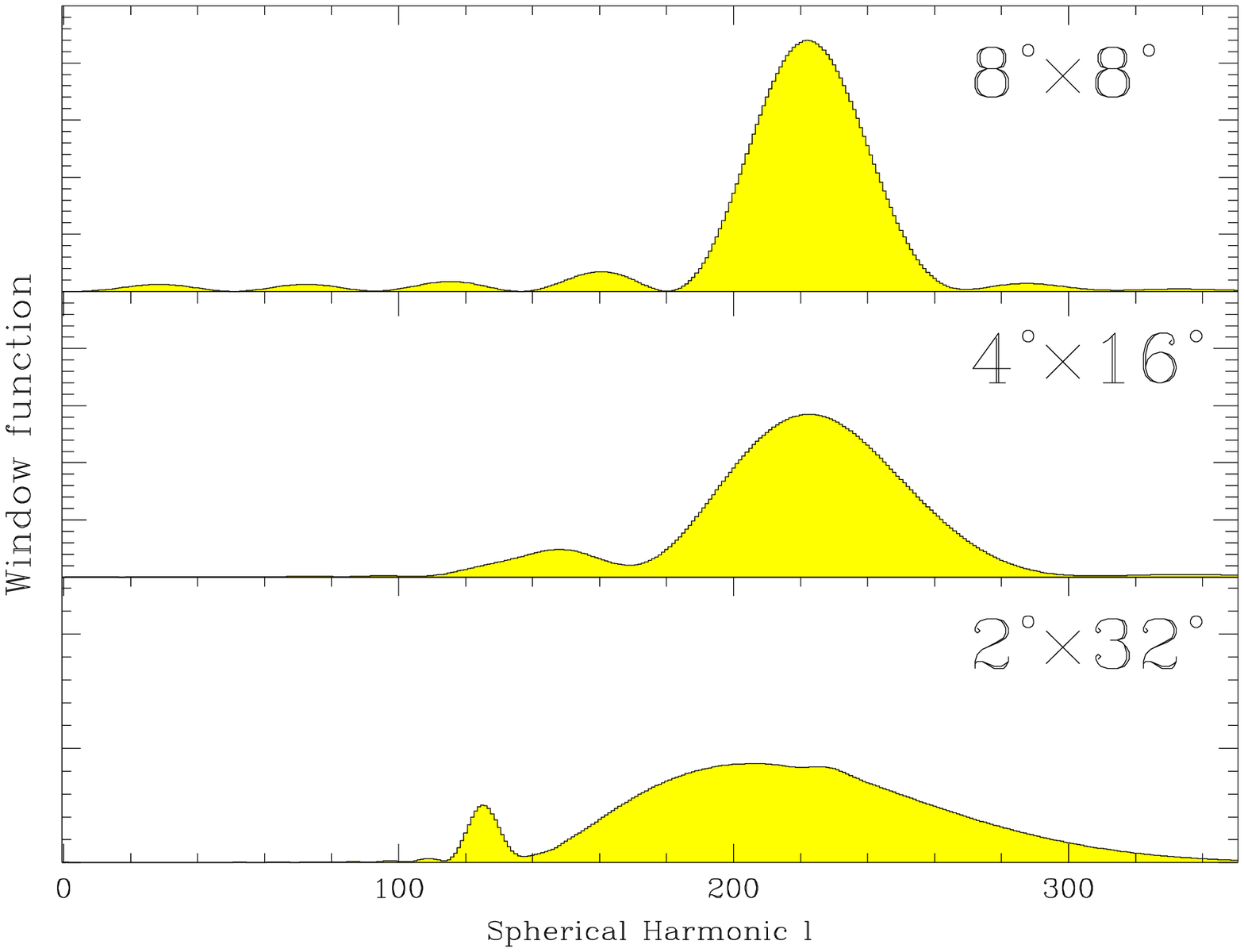}}}}\hfill
\vskip-12.3cm
\hglue0.8cm{{\vbox{\epsfxsize=2.5cm\epsfbox{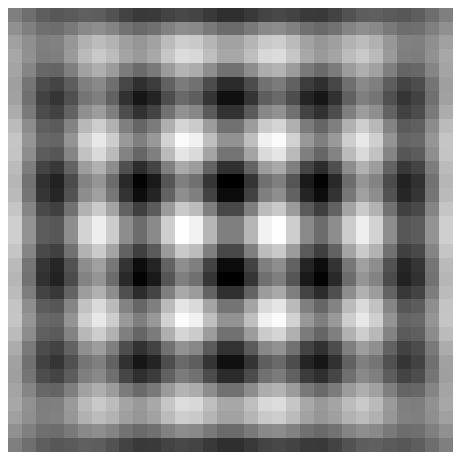}}}}\hfill
\vskip1.1cm
\hglue0.8cm{{\vbox{\epsfxsize=5cm\epsfbox{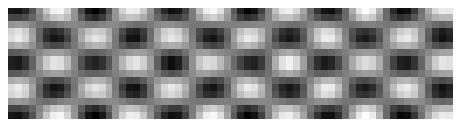}}}}\hfill
\vskip2.1cm
\hglue0.8cm{{\vbox{\epsfxsize=10cm\epsfbox{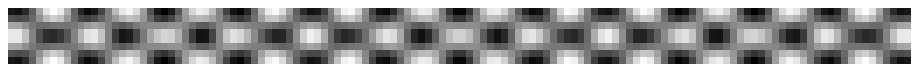}}}}
\vskip4.8cm
\caption{
Eigenmodes in real space and Fourier space.
The window functions are plotted for the signal-to-noise
eigenmodes with $\leff\sim 220$ for three rectangular sky patches
of the same area (64 square degrees), and are seen to
be wider for the skinnier patches.
The spatial eigenmodes themselves are also shown (inset).
}
\label{shapeFig}
\end{figure}

\begin{figure}[phbt]
\nothing
\vskip4.8cm
\hglue0.8cm{{\vbox{\epsfxsize=10cm\epsfbox{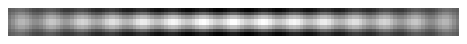}}}}\hfill
\vskip2.90cm
\hglue0.8cm{{\vbox{\epsfxsize=10cm\epsfbox{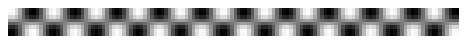}}}}\hfill
\vskip2.90cm
\hglue0.8cm{{\vbox{\epsfxsize=10cm\epsfbox{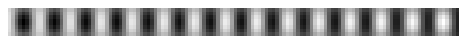}}}}
\vskip-12.0cm
{{\vbox{\epsfxsize=16cm\epsfbox{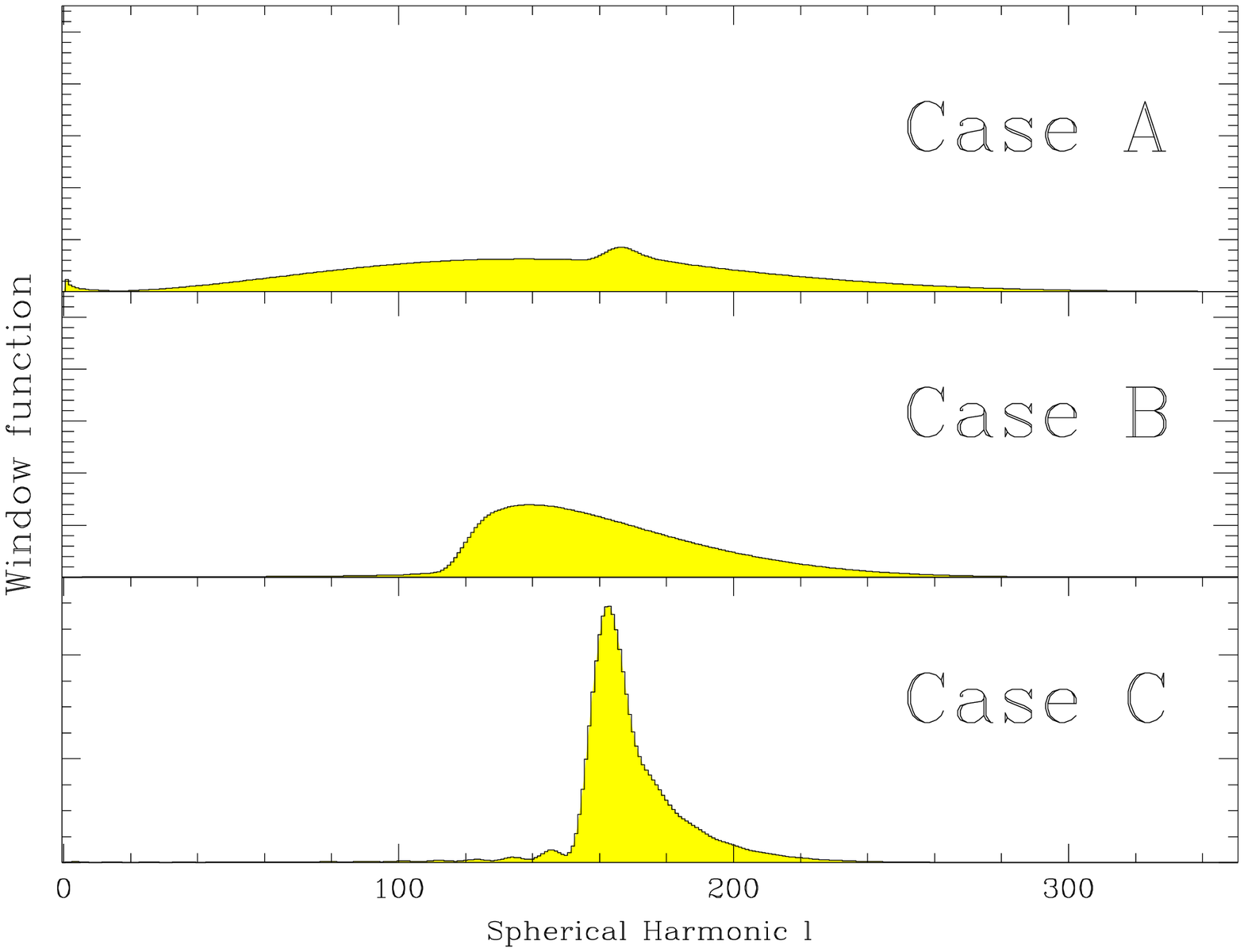}}}}\hfill
\caption{
The signal-to-noise eigenmodes 53, 52 and 51 (from top to bottom) and
the corresponding window functions are plotted for 
a $2^\circ\times 32^\circ$ rectangular sky patch.
}
\label{positionFig}
\end{figure}

\begin{figure}[phbt]
\centerline{{\vbox{\epsfxsize=16cm\epsfbox{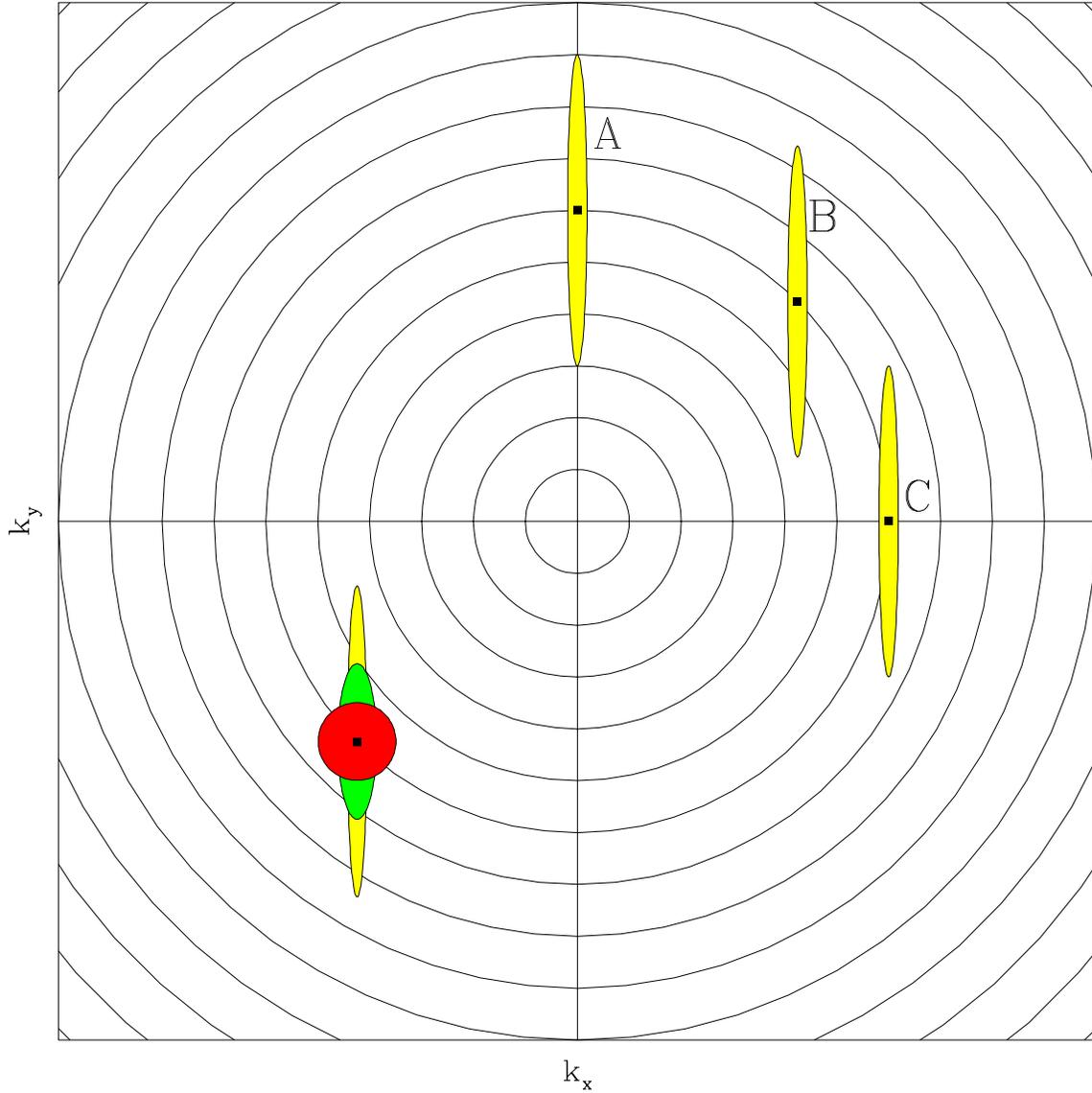}}}}
\caption{
Understanding window functions.
The two-dimensional Fourier transforms of the 
three eigenmodes from the previous 
figure are schematically illustrated by the ellipses at A, B and C.
The width of the one-dimensional window functions corresponds to their 
{\it radial} extent, {\it i.e.}, to how many of the circles they cross, so 
C gives a much narrower window than A in \fig{positionFig}.
The situation for the eigenmodes in \fig{shapeFig} is illustrated by the 
ellipses to the lower left.
}
\label{kspaceFig}
\end{figure}

\Fig{shapeFig} shows that more oblong regions generally produce
inferior (wider) window functions, but \fig{positionFig} illustrates
that there is also a strong dependence
on whether the oscillations are mainly in the narrow or wide direction.
All of this can be readily understood by considering 
two-dimensional rather than one-dimensional window functions, as illustrated
in \fig{kspaceFig}.
In the context of 3D galaxy redshift surveys, 
a mode probes  
a weighted average of the power in three-dimensional Fourier space,
and it is well-known that this 
weight function (3D window function) is simply the square modulus of
the Fourier transform of the mode itself.
The situation is analogous in the CMB case \cite{MH97}: in the flat sky approximation,
we can replace the $(\l,m)$ multipole space by a 2D Fourier space
$(k_x,k_y)$, and we can compute the 2D window function by simply 
Fourier transforming the signal-to-noise 
eigenmodes of \eq{RectEigenEq}, as illustrated in \fig{kspaceFig}.
The {\it shape} of the 2D window, schematically illustrated by the ellipses,
basically only depends on the shape of the
sky patch. It is of order $(\Delta k_x,\Delta k_y)\sim (\theta_x^{-1},\theta_y^{-1})$,
so since the $32^\circ\times 2^\circ$ map is 16 times wider than it is high, the
ellipses corresponding to its window functions are drawn 16 times higher than wide.
The central {\it location} of a window function is determined by
the wave vector $(k_x,k_y)$ in \eq{RectEigenEq}.
For instance, mode C in \fig{kspaceFig} has $k_y\sim 0$, {\ie},
virtually no vertical oscillations, so its window function  
lies straight to the right of the origin (the careful reader will
notice that since the eigenmodes contain $\cos(k_x x)$ rather than $\exp(ik_x x)$,
there should be a mirror image to the left, but this is omitted
to avoid cluttering up the figure).

The 1D window functions plotted in \fig{shapeFig} and \fig{positionFig}
depend only $\l$ (which corresponds to the radius $k$ in \fig{kspaceFig}), 
not on $m$ (roughly corresponding to the angular direction), and are 
essentially the angular average of the 2D window functions.
The width of the 1D window functions is therefore determined by
how many of the concentric circles are crossed in \fig{kspaceFig}:

\begin{itemize}

\item In \fig{positionFig}, mode A has the worst window function, 
because its longest extent is in the radial direction. Its spectral 
resolution is thus determined by $\theta_y$, the narrowest dimension of the patch. 

\item Mode C has the best window function, 
because its shortest extent is in the radial direction. 
In the limit $\leff\gg\Delta\l$ (where it is very far from the origin
in Fourier space),
its spectral resolution is thus determined by $\theta_x$, the broadest 
dimension of the patch. 

\item Modes like A and C constitute only a small minority, with 
typical modes being more like B, with comparable oscillations in 
the horizontal and vertical directions.  
For very oblong patches, the window function is a factor of $\sqrt{2}$
narrower for mode B than for mode A, {\ie}, it is still determined by the narrowest
direction alone.

\item The three cases compared in \fig{shapeFig} are all ``typical"
modes like B, but with patches whose aspect ratios
are 1, 4 and 16, respectively. These modes are illustrated to the lower
left in \fig{kspaceFig}, and show why the skinniest patch produces the worst 
result.

\end{itemize}

\subsection{Lesson 2: how to choose the map shape}

Above we saw how the window functions resulting from oblong sky patches 
could readily be understood in a two-dimensional Fourier space picture.
What does this tell us as regards the best choice of patch shape?

\subsubsection{What spectral resolution is needed?}

We want to be able to resolve all small-scale 
features in the power spectrum that carry
information about cosmological parameters. What is this scale?
Acoustic oscillations occur on a scale set by the horizon size at 
last scattering, corresponding to $\Delta\l\sim 200$ \cite{HSS97}
for an $\Omega=1$ CDM universe, and if $\Omega<1$, this scale $\Delta\l$
increases. To accurately measure the power spectrum, we clearly need
more than one measurement per Doppler peak, but a spectral resolution
$\Delta\l\sim 40$ would appear adequate for crude measurements, and 
$\Delta\l\sim 10$ should retain virtually all cosmological information.
The main exception is on the very largest scales, where the late
integrated Sachs-Wolfe effect can cause variations on a scale
$\Delta\l\sim 1$ if the universe has curvature of a non-zero cosmological 
constant. However, this is rather irrelevant to our present discussion,  
which is geared towards ground and balloon based experiments, 
since COBE has already measured these low multipoles with 
high spectral resolution, to near the cosmic-variance limit, and these
measurements are unlikely to be improved before the MAP mission flies.
Thus although some speculative models (\eg, \cite{Randall97})
introduce sharp features into the power spectrum, 
a spectral resolution $\Delta\l\sim 10-40$ appears sufficient for 
constraining the parameters of both standard 
inflationary and defect-based cosmologies.

\subsubsection{A rule of thumb}

Above we saw that the signal-to-noise eigenmodes
had $\Delta\l\sim 40$ for round regions of diameter $\sim 5^\circ$.
By eliminating ringing, the maximum-resolution method \cite{window} can reduce
this to $\Delta\l\sim 10-20$, but there are fewer uncorrelated modes 
that are this narrow, so to avoid increasing the sample variance, 
it is preferable to not go below this map size.
We also saw that the picture was more complex for oblong regions.
Although there are typically a small number of modes (which used alone
would give a large sample variance) with narrow window functions like
case C, the bulk of the modes have their width determined by the narrowest
direction of the survey. 
It is easy to see that a 
skinny mode centered at $(k_x,k_y)=k(\cos\varphi,\sin\varphi)$ will
have a window function width 
$\Delta\l\sim[\theta_x^{-2}\cos^2\varphi+\theta_y^{-2}\sin^2\varphi]^{1/2}$,
so if we use all the modes (which is necessary to attain the 
sample variance from Section~\ref{SizeSec}), then 
$(\Delta\l)^2$ gets averaged over $\varphi$ and we obtain
\beq{OblongDlEq}
\Delta\l\sim\sqrt{{\theta_x^{-2}+\theta_y^{-2}\over 2}}.
\eeq	
This means that if the patch is fairly oblong 
$(\theta_x\gg\theta_y)$, then $\theta_x$ becomes completely irrelevant
to the resolution, which is determined only by the narrowest 
direction, $\theta_y$.
Thus although it is possible to extract some useful constraints from
even narrower maps, we conclude our shape discussion with the following
rule of thumb:
\begin{itemize}
\item {\it Avoid maps that are skinnier than a few degrees in the narrowest direction.}
\end{itemize}
We have seen that as long as the narrowest direction $\simgt 5^\circ$, 
the situation is greatly simplifies, since {\it all} modes will be narrow enough in
Fourier space to be cosmologically useful. This means that one need not worry
about weeding out the widest modes, and can attain the minimal sample variance 
that the map area permits.

\section{FROM SCAN PATTERN TO MAP}

The previous section discussed measuring the power spectrum and 
cosmological parameters from a map, focusing on how to best chose its
size, shape, sensitivity and resolution. In this section, we turn 
to the preceding step in the data analysis pipeline: reducing time-ordered
data (TOD) to a map. Our focus will be on $1/f$ noise,
and how the power spectrum of the TOD noise in the time-domain
becomes processed into a map noise power spectrum in the multipole domain.
Specifically, once the shape and size of the map have been decided
as above, what is the best choice of scan pattern if we want
to minimize the map noise? To what extent is it worth complicating
the scan pattern to reduce the map noise?

\subsection{Mapmaking with 1/f-noise}

\subsubsection{The mapmaking problem}

The CMB mapmaking problem (see \cite{mapmaking} for a recent review)
is to estimate the map vector $\x$ of
the previous section from $M$ measured numbers $y_1,...,y_M$, 
which we will refer to as the {\it time-ordered data} (TOD),
and group into an $M$-dimensional vector $\y$.
Assuming that the TOD depends
linearly on the map, 
we can write 
\beq{LinearProblemEq}
\y = \A\x+\n
\eeq
for some known matrix $\A$ and some random noise vector $\n$.
Without loss of generality, we can take the noise vector to have zero
mean, {\ie}, $\expec{\n}=0$, so the noise covariance matrix is
\beq{NoiseCovEq}
\N\equiv\expec{\n\n^t}.
\eeq

\subsubsection{The solution}

All linear methods
can clearly be written in the form
\beq{WdefEq}
\xt = \W\y,
\eeq
where $\xt$ denotes the estimate of the map $\x$ 
and $\W$ is some $N\times M$ matrix that specifies
the method. 
If we make the choice
\beq{Method1Eq}
\W=[\A^t\M\A]^{-1}\A^t\M,
\eeq
where $\M$ is an arbitrary $M\times M$ matrix,
then $\W\A=\I$, which means that the reconstruction error $\err$,
defined as
\beq{errDefEq}
\err\equiv\xt-\x=[\W\A-\I]\x + \W\n
\eeq
is independent of $\x$. In other words, the
recovered map $\xt$ is simply the true map $\x$ plus some 
noise that is independent of the signal one is trying to measure.
We will therefore refer to $\err$ as the {\it noise map}, and study how
its statistical properties depend on the scan strategy
(specified by $\A$) and the detector noise characteristics 
(given by $\N$). Equations\eqno{WdefEq} and\eqno{errDefEq} show that
its covariance matrix $\NN\equiv \expec{\err\err^t}$
is given by 
\beq{SigmaDefEq}
\NN = \W\N\W^t = [\A^t\M\A]^{-1}[\A^t\M\N\M\A][\A^t\M\A]^{-1}
\eeq
if the matrix $\M$ is symmetric.

When chosing $\M$, it is clearly desirable to minimize 
the diagonal elements of $\NN$, 
the noise variance in the map, which gives
$\M=\N^{-1}$ and $\NN = [\A^t\M\A]^{-1}$.
However, noise correlations
manifested as off-diagonal elements in $\NN$ may also appear undesirable,
and one might fear that there is a tradeoff between these two evils
that muddles the issue as to how to choose $\M$. 
Fortunately, this is not the case.
$\M=\N^{-1}$ is the best possible method in the sense that the map
it produces can be shown \cite{mapmaking} to retain all the cosmological 
information from the TOD, even if the map is non-Gaussian,
and it has also been shown to be 
numerically feasible \cite{Wright96}, so there
is no need to settle for anything less.
If for instance Wiener-filtered or Maximum-Entropy filtered maps are desired,
these can always be computed directly from $\xt$ afterwards, 
without recourse to the TOD.

\subsubsection{Practical issues}

Although direct application of equations\eqno{WdefEq} and\eqno{SigmaDefEq} 
using a standard linear algebra package gives what we need
($\xt$ and $\NN$) in principle, this would be too slow to be useful 
in practice, since $\N$ is an $ M\times M$ matrix and $M$ is
typically between $10^6$ and $10^{10}$ \cite{Wright96}. 
Fortunately, this can be remedied by some numerical tricks.
A useful way of implementing the 
mapmaking algorithm described above 
was recently presented by Wright \cite{Wright96}.
It handles the inversion implicit in \eq{WdefEq} by solving 
for the vector $\xt$ iteratively, with the conjugate gradient method,
never computing $\NN$, 
which means that one avoids inverting large matrices explicitly. 
In the present paper, we specifically need the map noise covariance matrix 
$\NN$, to compute the noise power. Below we present tricks enabling
explicit calculation of $\NN$ for huge data sets
as long as the number of pixels $N\simlt 10^4-10^5$, 
which
should prove useful for some upcoming ground and balloon based CMB experiments.
The tricks make use of the fact that all three of the huge matrices
involved have very special properties: $\A$ is extremely sparse, 
and $\N$ and $\M$ can be replaced by matrices that are
both band-diagonal and circulant. Our treatment is slightly more general than 
Wright's Fourier approach \cite{Wright96} in that it 
treats discreteness and edge effects exactly and is applicable also 
if data blocks are too short to 
allow one to pre-whiten the noise exactly.

\subsubsection{The circulant matrix trick}

As this and the subsequent section are rather technical, the
reader not interested in data analysis {\it per se} is encouraged to 
jump directly to Section~\ref{FourScanSec}.

A square matrix $\C$ is said to be {\it circulant} \cite{Davis79} if 
each of its rows is merely the one above it cyclically shifted one notch
to the right, {\ie}, if $\C_{i+1,j+1}=\C_{ij}$, understood (mod $M$) for 
an $M\times M$ matrix. As we will return to below, circulant matrices 
have the useful property of being extremely fast to invert and multiply.

Assuming that the statistical properties of the detector noise are 
independent of time, the correlation between the noise $n(t)$ at two 
different times will depend only on the time separation:
$\expec{n(t)n(t')} = c(t-t')$ for some 
{\it time correlation function} $c$ (which is by definition symmetric;
$c(-\tau)=c(\tau)$).
Assuming that the measurements in the TOD are made at a
uniform rate in time, separated by some constant time interval
$\Delta t$ and starting at some time $t_0$, 
the noise covariance matrix $N$ thus takes the form
\beq{NoiseCorrEq}
\N_{ij} = \expec{n(t_0+i\Delta t) n(t_0+j\Delta t)} = c(|i-j|\Delta t).
\eeq
For instance, the $M=5$ case can be written
\beq{N5eq}
\N = 
\left(\begin{tabular}{ccccc}
$c_0$&$c_1$&$c_2$&$c_3$&$c_4$\\
$c_1$&$c_0$&$c_1$&$c_2$&$c_3$\\
$c_2$&$c_1$&$c_0$&$c_1$&$c_2$\\
$c_3$&$c_2$&$c_1$&$c_0$&$c_1$\\
$c_4$&$c_3$&$c_2$&$c_1$&$c_0$
\end{tabular}\right),
\eeq
where we have defined the noise correlations
\beq{cnDefEq}
c_n\equiv c(n\Delta t).
\eeq
It would be numerically useful if this were a symmetric circulant matrix. 
However, the above definition shows that the $M=5$ symmetric
circulant matrix takes the form
\beq{N5cEq}
\N_c = 
\left(\begin{tabular}{ccccc}
$c_0$&$c_1$&$c_2$&$c_2$&$c_1$\\
$c_1$&$c_0$&$c_1$&$c_2$&$c_2$\\
$c_2$&$c_1$&$c_0$&$c_1$&$c_2$\\
$c_2$&$c_2$&$c_1$&$c_0$&$c_1$\\
$c_1$&$c_2$&$c_2$&$c_1$&$c_0$
\end{tabular}\right).
\eeq
In other words, the requirement that it wraps around modulo $M$
specifies the upper right and lower left corners, requiring 
that $c_4=c_1$ and $c_2=c_3$, which would correspond to the correlation 
between the first and last observation equaling that between the 
first and second one, {\etc}
However, it is useful to decompose $\N$ as a sum of a circulant and a non-circulant
matrix, as 
\beq{CyclicDecompositionEq}
\N = \N_c + \N_s,
\eeq
where for our $M=5$ example, the latter is given by 
\beq{N5sEq}
\N_s = 
\left(\begin{tabular}{ccccc}
$0$&$0$&$0$&$c_3-c_2\>$&$\>c_4-c_1$\\
$0$&$0$&$0$&$0$&$\>c_3-c_2$\\
$0$&$0$&$0$&$0$&$0$\\
$c_3-c_2\>$&$0$&$0$&$0$&$0$\\
$c_4-c_1\>$&$\>c_3-c_2$&$0$&$0$&$0$
\end{tabular}\right).
\eeq
The subscript $s$ denotes {\it sparse}, since as we will see in the
next section, we can make the correlations $c_n$ vanish for $n\gg 1$. 
If there is some integer $L\ll M$ such that $c_n=0$ for $n>L$, 
then $\N_s$ will contain merely $L(L+1)$ non-zero elements, and it will be
trivial to multiply by (which as we will see is all we need to do with it). 
The circulant matrix $N_c$ will be band-diagonal and contain 
$(2L+1)M$ nonzero numbers, {\ie}, a factor $\sim 2N/L\gg 1$ more
than $\N_s$. For typical applications, $L\sim 10-100$ and $M\sim 10^6-10^{10}$, 
so when performing matrix operations with $\N$, the $\N_c$-term will 
completely dominate over the $\N_s$-term. Specifically, $\N^{-1}\approx \N_c^{-1}$.
We now come to our first speed trick. Our mapmaking algorithm computes the correct
$\NN$ for the resulting map $\xt$ for {\it any} choice
of $\M$. Minimizing the map noise variance gave $\M=\N^{-1}$,
so this variance will clearly increase only to second order if we change
$\M$ slightly. Let us take advantage of this by replacing the strictly optimal
choice $\M=\N^{-1}$ by the more convenient choice
\beq{ConvenientMeq}
\M=\N_c^{-1}. 
\eeq

To proceed, we need to be able invert the $M\times M$ matrix $\M$.
Being able to compute $\N_c^{1/2}$ is also useful at times, 
since it enables one to
make Monte Carlo  simulations of the noise
using the equation $\n=\N^{1/2}\z\approx\N_c^{1/2}\z$, 
where $\z$ is a vector of uncorrelated normalized Gaussian random variables.
We now describe how to do both.
The action of {\it any} function on a
symmetric matrix is defined as the corresponding real-valued
function acting on its eigenvalues:
Since all symmetric matrices $\C$ can be diagonalized as
\beq{CdiagEq}
\C=\R\LL\R^t,
\eeq
where $\R$ is orthogonal ($\R\R^t=\I$) and $\LL = \diag{\lambda_i}$ 
is diagonal and real, one can extend any mapping $f$
on the real line to symmetric matrices by defining
\beq{MatricFuncDefEq}
f(\R\diag{\lambda_i}\R^t) \equiv \R\diag{f(\lambda_i)}\R^t,
\eeq
or more explicitly, 
\beq{MatricFuncEq2}
f(\C)_{mn} = \sum_k\R_{mk}\R_{nk} f(\lambda_k).
\eeq
It is easy to see that this definition is consistent with
power series expansions whenever the latter converge.

Circulant matrices have the great
advantage that they all commute. This is because 
they can all be diagonalized by the same matrix $\R$, an
orthogonal version of the discrete Fourier matrix.
If $\C$ is
symmetric, positive-definite, circulant and infinite-dimensional
(the latter is an excellent approximation as long as $M\gg L$),
then \eq{MatricFuncEq2} simplifies to \cite{FKM}
\beq{FKMeq}
f(\C)_{mn} = 
{1\over 2\pi}\int_{-\pi}^{\pi} f[\lambda(\varphi)]cos[(m-n)\varphi] d\varphi,
\eeq
where $\lambda(\varphi)$, the {\it spectral function} of the matrix,
is the function whose Fourier coefficients are
row zero of $\C$, {\ie}, 
\beq{SpectralFuncDefEq}
\lambda(\varphi) = \sum_{n=-\infty}^{\infty} c_n e^{in\varphi}.
\eeq
Note that $f(\C)$ is circulant as well. In particular, 
the inverse $\N_c^{-1}$, which we can compute by chosing $f(x)=1/x$ in 
\eq{FKMeq}, will also be circulant.
It is easy to see that multiplying two circulant matrices 
also produces a circulant matrix, and that this corresponds to multiplying 
their spectral functions. 
This is equivalent to convolving their $0^{th}$ 
rows, which is also extremely quick 
if both matrices are band-diagonal. 
Thus all the operations on circulant matrices in \eq{SigmaDefEq} 
(inverting $\N_c$ to obtain $\M$, multiplying $\M$ with $\N_c$, {\etc}), 
produce new circulant matrices, so all we ever need to store is 
row zero of each square matrix being manipulated. 

What about the matrix $\A$? For a single-horned experiment,
all its entries are zero except that there
is a single ``1" on each row. Letting $N_i$ denote the number of the
pixel pointed to at the $i^{th}$ observation (at time
$t=t_0+i\Delta t$), we have $\A_{ij}=1$ if 
$N_i=j$, $\A_{ij}=0$ otherwise.
This makes it very simple to multiply by both
$\A$ and $\A^t$. 
For instance, for any vector $\a$, we can compute the vector 
$\b\equiv\A^t\a$ by a single loop over $i=1,...,M$ \cite{Wright96}:
$b(N_j) := b(N_j) + a(i)$,
simply summing the temperature measurements of each pixel.
Multiple beams introduce virtually no additional difficulty.
For double-horned experiments like COBE and MAP, there
are simply two non-zero entries in each row, 
a ``1" and a ``$-$1". 

\subsubsection{A trick for making the matrices band-diagonal}

The computation of the matrices $[\A^t\M\A]$ and 
$[\A^t\M\N\M\A]$ can be further accelerated by making the circulant 
matrices $\N_c$ and $\M$ band-diagonal.

\paragraph{The noise model:}

The noise correlations $c_n$ from \eq{cnDefEq} 
can be computed as
\beq{cnCalcEq}
c_n = {1\over\pi}\int_0^{\infty} P(\omega)\cos(n\omega\Delta t)d\omega,
\eeq
where the noise (time) power spectrum $P(\omega)$ is simply the Fourier transform
of the time correlation function $c(\tau)$.
The noise characteristics of most CMB detectors can be well fit by 
an expression of the form:
\beq{NoisePowerModelEq}
P(\omega) 
= \sigma^2\left[1+{\omega_k\over\omega}+
{\left(\omega_b\over\omega\right)^2}\right]|\Wh(\omega)|^2.
\eeq
The three terms in square brackets correspond to white noise, 
$1/f$ noise and so-called brown noise, respectively, and the ``knee"
frequencies $\omega_k$ and $\omega_b$ determine where they yield 
the same power as the white noise.
(The angular frequency $\omega$ is related to the frequency $f$ by $\omega=2\pi f$.)
Most CMB detectors have no brown noise component ($\omega_b=0$) --- we
are including it here for pedagogical reasons, since it turns out to be 
very simple to understand its effects, and the properties of $1/f$ noise 
are intermediate between the simple white and brown cases.
$W$ is a window function specifying what kind of analog
smoothing (convolution) was performed on the time signal before sampling it.
Here we will follow \cite{Janssen96} by assuming ``boxcar" smoothing
where $y_i$ is the average of the signal measured during a 
time interval $\Delta t$. This corresponds to 
$W(\tau)=\theta(\Delta t/2-|\tau|)/\Delta t$. Fourier transforming this gives
\beq{BoxcharWhEq}
\Wh(\omega) = j_0(\omega\Delta t/2),
\eeq
where $j_0(x)=\sin x/x$.

Substituting \eq{cnCalcEq} into \eq{SpectralFuncDefEq}, 
we see that the relation between the power spectrum and the spectral function is 
\beq{SpectralFuncCalcEq}
\lambda(\varphi) = \sum_{n=-\infty}^{\infty} P[(\varphi+2\pi n)/\Delta t], 
\eeq
where $P(-\omega)=P(\omega)$. In other words, 
the power spectrum simply ``wraps around" onto itself many times, 
with all power above the Nyquist sampling 
frequency $\pi/\Delta t$ getting aliased down to lower frequencies.

\paragraph{The white noise case:}

White noise alone ($\omega_k=\omega_b=0$) gives the 
trivial case of uncorrelated noise: $c_n$ vanishes except for 
$n=0$, so $\N\propto\I$, $\M\propto\I$, and the mapmaking reduces to simply
averaging measurements of each pixel in the map.  
The variance $\NN_{ii}$ in each map pixel is simply $\sigma^2$ 
divided by the number of times it was observed, so if the sky patch has been 
covered uniformly, we obtain the familiar case $\NN\propto\I$
corresponding to white noise in the map, whose angular power spectrum 
is given by \eq{ClNoiseEq}.

\paragraph{The correlated noise case:}

If $1/f$ noise or brown noise is present, then the integral
in \eq{cnCalcEq} diverges at low frequencies.
This means that slow drifts will completely dominate the 
noise, and that all the coefficients $c_n$ will be equal
(the noise at any two times will be perfectly correlated).
This is of course not a problem in practice, since we can remove 
these slowly varying offsets -- it is merely a numerical nuisance,
and is easily eliminated by replacing the TOD $\y$ by a high-pass 
filtered data set
\beq{ytDefEq}
\yt\equiv\D\y,
\eeq
where $\D$ is some appropriate circulant matrix.
The new noise covariance matrix becomes
\beq{NtDefEq}
\Nt\equiv \expec{(\D\n)(\D\n)^t}=\D\N\D^t.
\eeq
Using $\yt$ instead if $\y$ as the starting point for the mapmaking 
process, \eq{LinearProblemEq} becomes $\yt=\At\x+\nt$, where $\At\equiv\D\A$,
so equations\eqno{WdefEq} and\eqno{SigmaDefEq} follow with tildes 
on all matrices, or explicitly, eliminating
all tildes, 
\beqa{Weq2}
\NN&=&[\A^t\D^t\M\D\A]^{-1} 
[\A^t\D^t\M\Nt\M\D\A]
[\A^t\D^t\M\D\A]^{-1},\nonumber\\
\W&=&[\A^t\D^t\M\D\A]^{-1}\A^t\D^t\M\D
\eeqa
where $\Nt=\Nt_c+\Nt_s$ as before and $\M=\Nt_c^{-1}$.

\begin{figure}[phbt]
\centerline{{\vbox{\epsfxsize=16cm\epsfbox{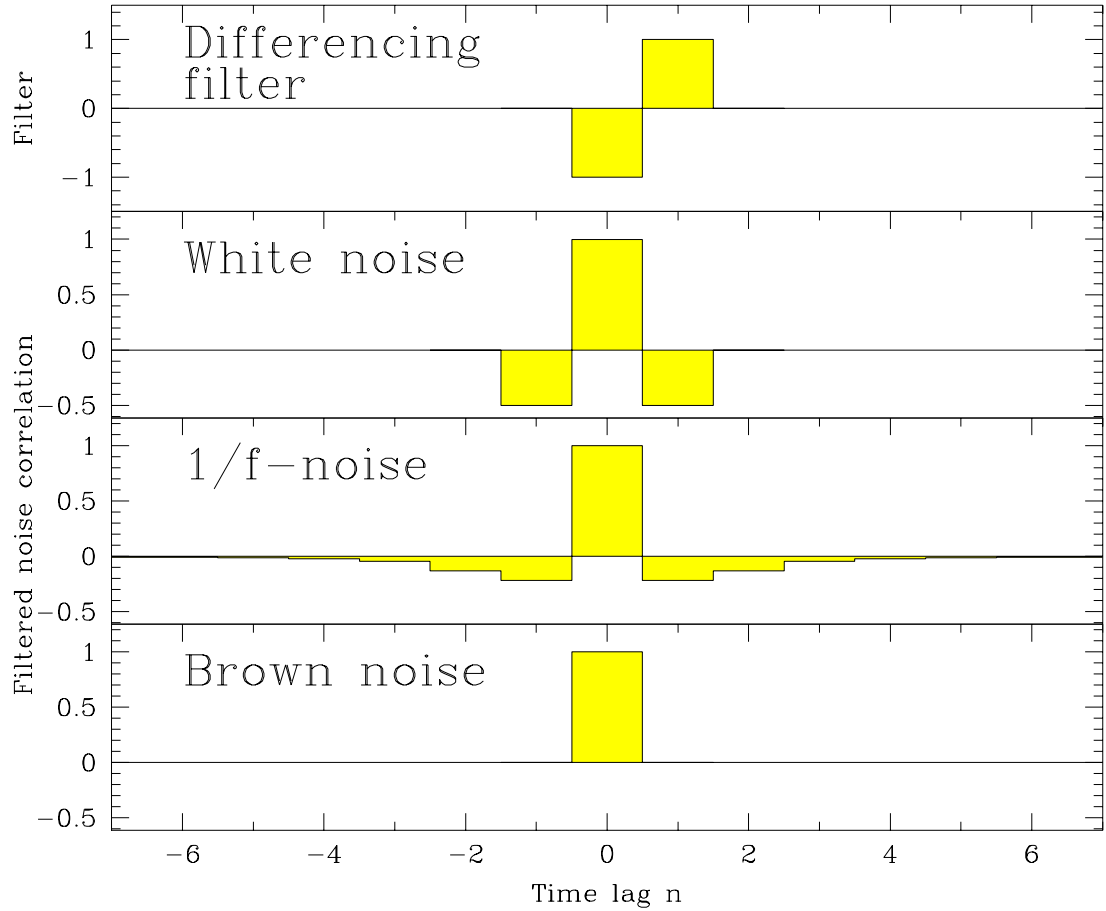}}}}
\caption{
When each measurement is subtracted from the one following it
(using the differencing filter in the top panel), the correlation 
functions resulting from white, $1/f$ and brown noise take the form
shown in the three lower panels.
}
\label{NoiseCorrFig1}
\end{figure}

\Fig{NoiseCorrFig1} shows the effect of the simple choice
where all components of $\D$ vanish except $\D_{ii} = -1$
and $\D_{i,i+1} = 1$.
This corresponds to
simply taking differences of consecutive observations: 
${\tilde y}_i = y_{i+1}-y_i$, and row zero of $D$ (the convolution filter)
is plotted in the top panel. 
The bottom three panels show that 
whereas $\N$ was pathological with non-zero and constant 
correlations extenting arbitrarily far from the diagonal, 
$\Nt$ is almost diagonal.
These correlation functions were computed
as follows.
\Eq{SpectralFuncDefEq} shows that the spectral function of $\D$
is $\lambda(\varphi)=e^{i\varphi}-1$, so that of 
the matrix $\D \D^t$ is $|e^{i\varphi}-1|^2=4\sin^2(\varphi/2)$.
$\Nt=\D\N\D^t\approx\D\N_c\D^t = \N_c\D\D^t$ can therefore be computed
explicitly by combining equations\eqno{FKMeq} 
and\eqno{SpectralFuncCalcEq}, which gives
\beq{cnIntegralEq}
c_n \propto \int_0^{\infty} 
j_0^2\left({x\over 2}\right)\sin^2\left({x\over 2}\right) x^\alpha\cos(nx) dx,
\eeq
where $\alpha=$0, -1 and -2 corresponds to white, 1/f and brown noise, respectively.
Performing the integral for these three cases gives
\beq{DNDcasesEq}
c_n \propto \left\{
\begin{tabular}{ll}
$2\delta_{0n}-\delta_{1|n|}$&for white noise,\\
$\phi(n)$&for 1/f noise,\\
$\delta_{0n}$&for brown noise,\\
\end{tabular}
\right.
\eeq
where the function $\phi$ is given by
\beqa{phiDefEq}
\phi(n)&=&(n-2)^2\ln|n-2| - 4(n-1)^2\ln|n-1| +6n^2\ln|n|\nonumber\\
&-&4(n+1)^2\ln|n+1| + (n+2)^2\ln|n+2|,
\eeqa
and $0\ln 0$ is to be interpreted as 0.
For $n>2$, this is accurately approximated by
\beq{phiApproxEx}
\phi(n) \approx -\left[{2\over n^2}+{2\over n^4}+{3\over n^6}\right],
\eeq
which shows that even the $1/f$ noise, which produces the widest correlation function of the
three types, is roughly band-diagonal and can be safely truncated at say $n>L\sim 100$. 

As the figure shows, brown noise has the property that all the differences are uncorrelated.
Thus the noise $n(t)$ exhibits Brownian motion over time, 
which explains its nickname.
Brown noise drifts like $t^{1/2}$ over time, whereas $1/f$ noise 
is much milder in
that it drifts only logarithmically in $t$.

When faced with a noise time stream $\n$ from real data, a good way to diagnose is is to 
compute the differences ${\tilde n}_i=n_{i+1}-n_i$ and 
estimate $c_k$ as the time average of the product ${\tilde n}_i{\tilde n}_{i+k}$.
Fitting this with a linear combination of the three templates in 
\fig{NoiseCorrFig1} will indicate the level 
at which the three basic types of noise are present, although for a more accurate
model, it is better to compute the noise spectral function 
directly by substituting the measured $c_n$-coefficients into 
\eq{SpectralFuncDefEq}.

\paragraph{Pre-whitening}

\begin{figure}[phbt]
\centerline{{\vbox{\epsfxsize=16cm\epsfbox{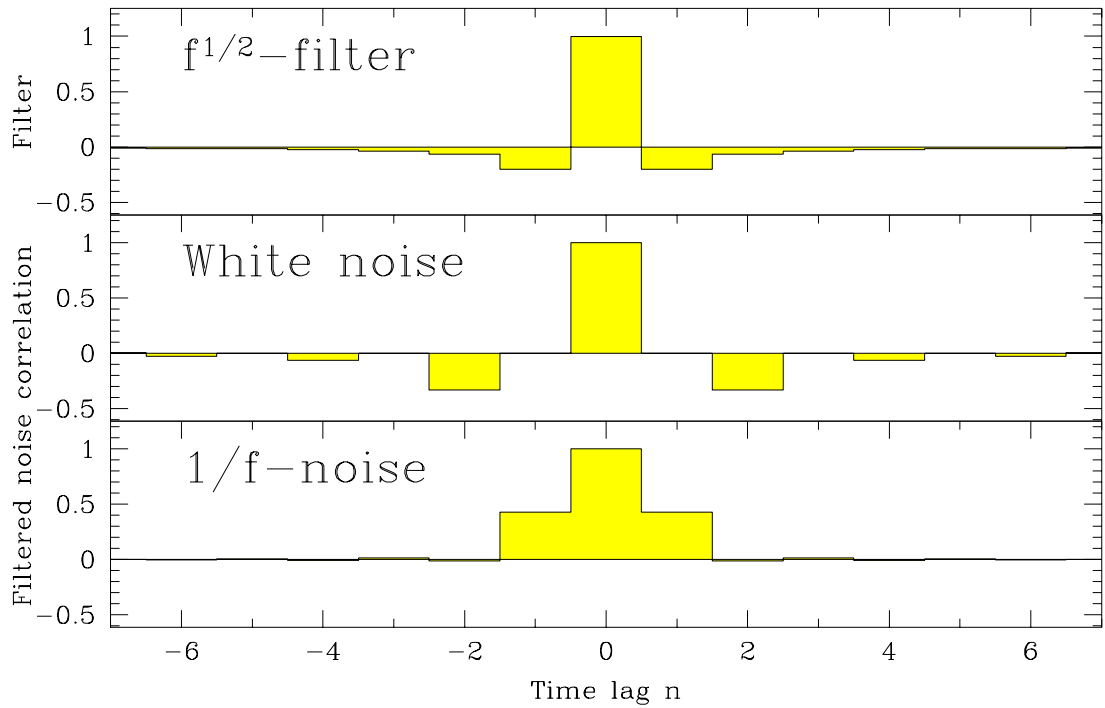}}}}
\caption{
When the time stream is convolved with the ``half differencing'' 
filter in the top panel, the correlation functions resulting from 
white and $1/f$ noise are as shown in the lower panels.
As opposed to in the previous figure, the $1/f$ correlation function
does not sum to zero, which makes $\M$ band-diagonal.
}
\label{NoiseCorrFig2}
\end{figure}

\begin{figure}[phbt]
\centerline{{\vbox{\epsfxsize=16cm\epsfbox{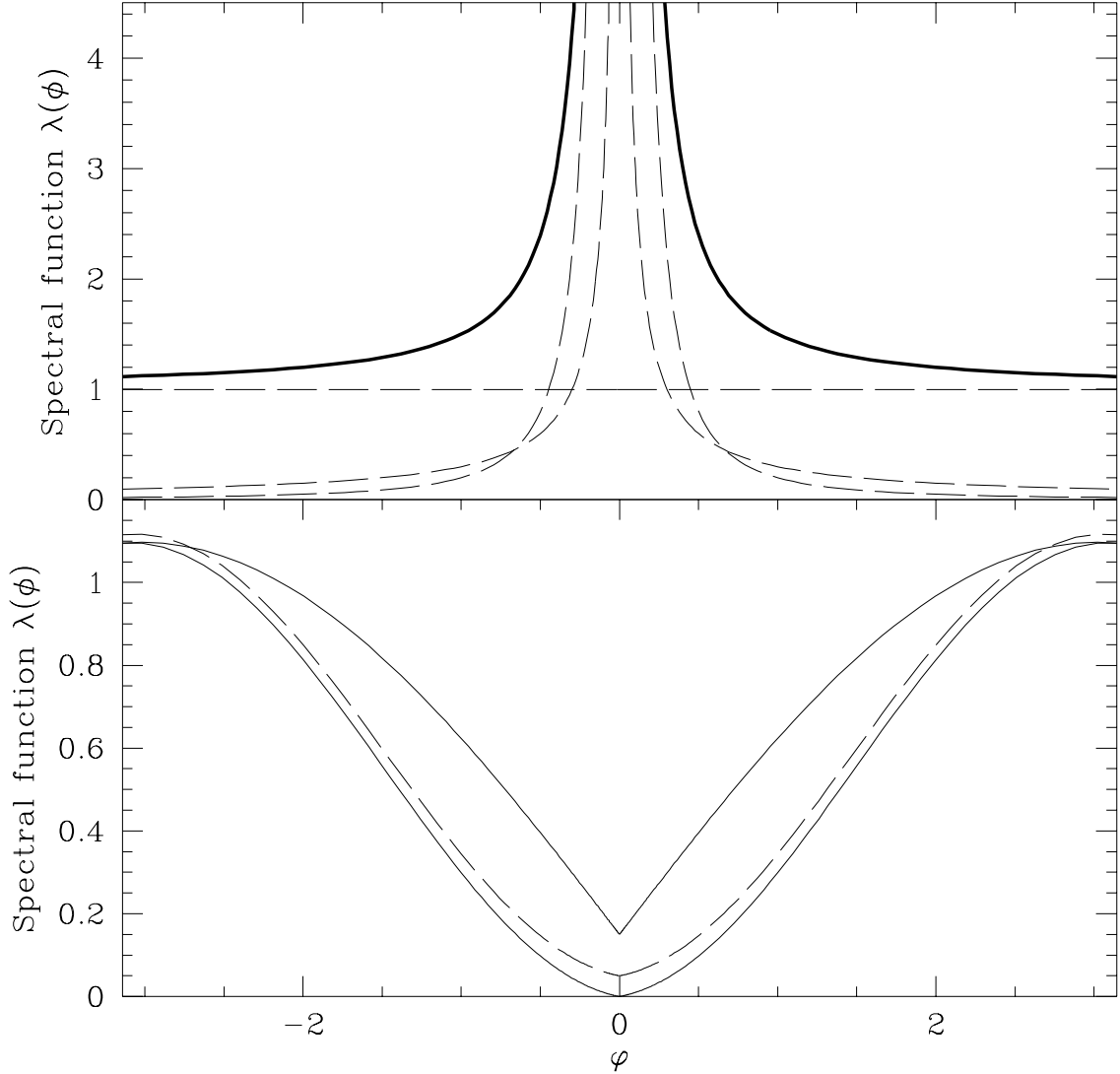}}}}
\caption{
Spectral functions. 
The top panel shows the spectral function for a sample noise covariance matrix (solid line)
and its decomposition into white, 1/f and brown noise  (dashed curves). 
The bottom panel shows the same spectral function after differencing the data (dashed curve), 
corresponding to multiplication with $\sin^2(\varphi/2)$. In the absence of brown noise
(lower solid curve), $\lambda(0)=0$ which is inconvenient for computing $\N^{-1}$, but this
problem can be eliminated by using different high-pass filter -- the upper solid curve
differs by a factor $|\sin(\varphi/2)|$.
}
\label{SpectralFuncFig}
\end{figure}

To be able to make maps with \eq{Weq2}, we want all the circulant matrices that appear to be 
close to diagonal. We saw that when $\D$ is simply the differencing matrix, 
$\D$ and $\Nt_c$ are indeed band-diagonal. But what about $\M$, the inverse of $\Nt_c$? 
\Fig{SpectralFuncFig} shows the spectral function $\lambda(\varphi)$ of $\N_c$ and $\Nt_c$ 
when all three types of noise are present. 
For this case, $\lambda(\varphi)>0$ for all $\varphi$, so its inverse, which is the spectral
function of $\M$, will be smooth and well-behaved, giving a band-diagonal $\M$ as desired.
If there is no brown noise component, however, we will have $\lambda(\varphi)\propto |\varphi|$
for $|\varphi|\approx 0$ (lower panel), so the spectral function of $\M$ blows up near the
origin and $\M$ will have inconvenient non-zero elements arbitrarily far from the diagonal.
(Differencing multiplies by $f^2$ near the origin, so this ``overkill'' of $1/f$ noise
produces an $\M$-matrix with $1/f$ noise.)
This can be remedied by a better choice of $\D$, whose spectral function exactly neutralizes the 
1/f-noise at the origin. A simple choice that does this is the $\D$ whose spectral function
is $|\sin(\varphi/2)|^{1/2}$, and is a ``half-difference"
in the sense that doing it four times is equivalent to double differencing, which we saw 
multiplied the spectral function by $\sin^2(\varphi/2)$.
The explicit convolution filter is plotted in \fig{NoiseCorrFig2}, and is seen to keep both 
white and $1/f$ noise close to diagonal. 
Another attractive option \cite{Wright96} 
is to {\it prewhiten} the data, by chosing the high-pass filter 
$\D$ to have a spectral function that is the inverse square root of the spectral function
of $\N_c$. This reduces $\N_c$ (and hence also $\M$) to the identity matrix, 
so $\D^t\D$ is the only circulant matrix remaining in \eq{Weq2}.
We remind the reader that all choices of $\D$ produce the exact same answer, 
so the choice is merely one of numerical convenience. The 
choice of $\D$ makes very little difference in practice as well,
as long as one ensures that 
the resulting spectral function of $\Nt_c$ is smooth and non-zero, since multiplying the 
various circulant matrices together in \eq{Weq2} is virtually instantaneous compared to
the other numerical steps. 

\subsection{Case study 3: four scan patterns}

\label{FourScanSec}

\begin{figure}[phbt]
\centerline{{\vbox{\epsfxsize=16cm\epsfbox{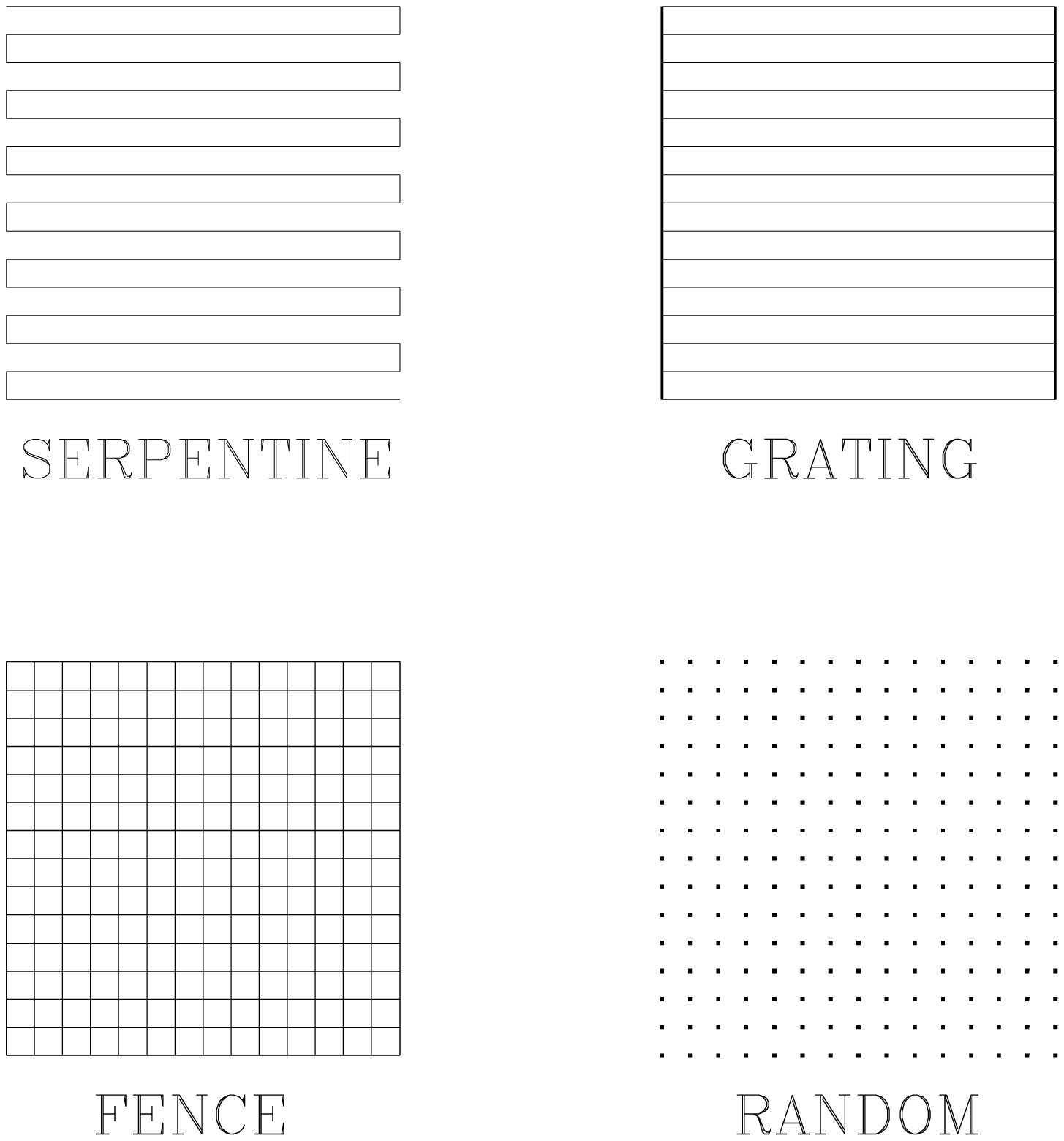}}}}
\caption{
Schematic illustration of the four scan patterns described in the text.
}
\label{ScanPatternFig}
\end{figure}

Let us consider a square sky patch of diameter $8^\circ$, divided into 
$N=32\times 32$ pixels, scanned
in four different ways as illustrated in \fig{ScanPatternFig}:
\begin{enumerate}
\item {\bf Serpentine scan:} the beam sweeps back and forth horizontally,
gradually shifting downward, not crossing its path until the entire patch has
been covered. This is reversed, then everything is repeated.
\item {\bf Grating scan:} a serpentine scan is augmented with an equal amount of
time spent scanning up and down along the left and right edges.
\item {\bf Fence scan:} two sets of serpentine scans are performed in succession, 
one horizontal and one vertical (rotated by $90^\circ$).
\item {\bf Random scan:} The beam jumps to a random pixel after each observation,
but in such a way that all pixel pairs are connected equally many times. 
\end{enumerate}
In all cases, we make $M=2^{21}\sim 2\times 10^6$ observations.
These simple scanning strategies span the entire range of ``connectedness''
available in real-world experiments, with the serpentine scan being the least 
connected one possible and the random scan at the other extreme.
An experiments with disjoint strips such as Tenerife is more
similar to the serpentine case, whereas 
double-beam differencing experiments such as 
COBE are very well connected and more similar
to the random case.
A {\it Planck} scan strategy pattern with great circles (pointing $90^\circ$
away from the spin axis) would be reminiscent of the grating case, with disjoint
strips (in this case circular arcs) connected together at two points (at the poles,
where they circles). 
Several recently flown and proposed balloon 
experiments have linear or circular scans intersecting at a variety of
angles, which makes them similar to the fence case.
If {\it Planck} points $70^\circ$ away from the spin axis, as originally proposed, 
its scan pattern would also be rather fence-like.

To be able to isolate how the features of these scan patterns affect the
ability to minimize various types of noise, let us first study the
effect of white and $1/f$ noise separately, then compute some cases where they
are present in combination.

\subsection{Measuring the noise power spectrum}

\label{NoisePowerSec}

We first note that pixel noise strictly speaking 
does not have a power spectrum at all in general,
since its statistical properties are not isotropic.
Rather, the quantity of interest is how much power it adds to our
estimates of the CMB power spectrum coefficients (the expectation value
of this noise contribution is of course subtracted out to make the 
$C_\l$-estimates unbiased, but the noise still contributes to the error bars
on these estimates). 
We will therefore compute the noise power just as we would compute the CMB
power, using the minimum-variance method \cite{cl}.
Using a simple white noise prior, this corresponds to computing
\beq{ClNoiseCalcEq}
\Cnoise_\l = {\tr\P^\l\NN\over\tr\P_\l\C},
\eeq
where the matrix $\P^\l$ is given by
\beq{PdefEq}
\P^\l_{ij} = P_\l(\rh_i\cdot\rh_j)
\eeq
and the $P_\l$ are Legendre polynomials.
$\C$ is the covariance matrix that would result from 
a white noise power spectrum; $\C=\sum_\l (2\l+1)\P^\l B_\l^2/4\pi$.

\subsection{Pure white noise}

\begin{figure}[phbt]
\centerline{{\vbox{\epsfxsize=16cm\epsfbox{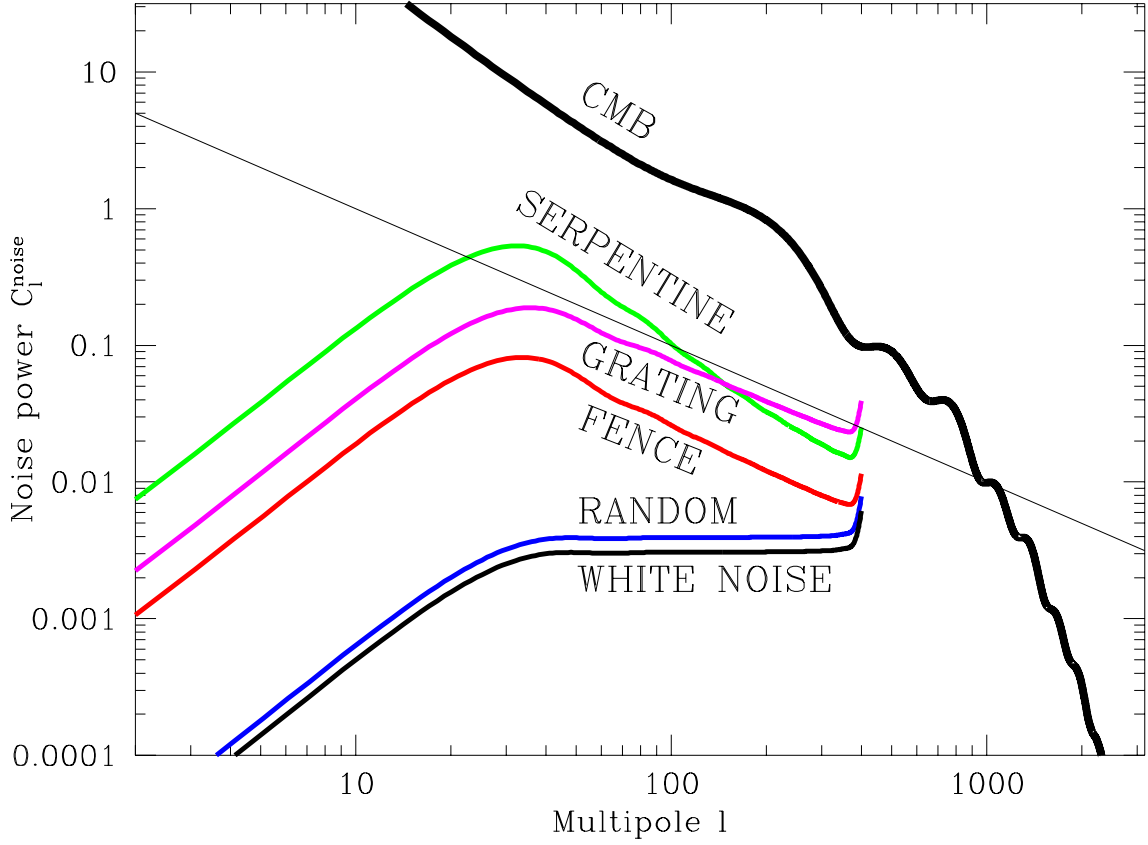}}}}
\caption{
The noise power spectrum $\Cnoise_\l$ is plotted for our four scan patterns
given pure $1/f$ noise. The corresponding noise power for white detector noise
(which is identical for the serpentine, fence and random scans) is plotted
below for comparison, as well as  
a standard CDM power spectrum (top). The straight line has slope $\l^{-1}$,
just like the serpentine and fence power spectra.
}
\label{NoisePowerFig1}
\end{figure}

Our four scan patterns were chosen such that all pixels are observed
the same number of times (except for the pixels on the left and
right edges of the grating scan). 
This means that if only white noise is present,
we will have uniform uncorrelated map noise ($\NN\propto\I$), and the simple
expression in \eq{ClNoiseEq} applies.
This is the lowermost line plotted in \fig{NoisePowerFig1}.
To draw attention to the simple shapes of the noise power spectra,
we are not including the beam smearing effect here ($B_\l=1$), which would
otherwise make $\Cnoise_\l$ blow up exponentially for large $\l$.
Note that the curve is only horizontal (as predicted by \eq{ClNoiseEq})
on the scales probed by the experiment.
On scales comparable to the pixel separation, artifacts appear (this is irrelevant
when the map is properly oversampled, since beam smoothing destroys any CMB signal on
these scales). On scales larger than the patch size 
(corresponding to $\l\simlt 30$), the noise power drops as $\l^2$
since the mapmaking algorithm is insensitive to the monopole (mean) of the
map -- this occurs automatically when $1/f$ noise is present, as
the method removes baseline drifts. (The matrix
$[\A^t\D^t\M\D\A]$ to be inverted in \eq{Weq2} will have one vanishing eigenvalue,
corresponding to the mean,
which is dealt with using the pseudo-inverse approach described 
in the appendix of \cite{cl}).

\subsection{Pure $1/f$ noise}

The other curves in \fig{NoisePowerFig1} show the effect of pure 1/f-noise.
Note that there is no scale in the problem other than the patch size
(to the left of which the monopole removal starts suppressing the power)
and the pixel separation scale (where irrelevant artifacts appear and we
have truncated the curves), so it should come as no surprise that the curves
are rather featureless between these two scales. (The $1/f$ knee frequency 
cannot imprint a feature here, since it is of course only defined when 
there is white noise present.)
The normalization is arbitrary -- doubling the receiver noise 
merely doubles the power spectrum.

\begin{figure}[phbt]
\centerline{\rotate[r]{\vbox{\epsfxsize=14cm\epsfbox{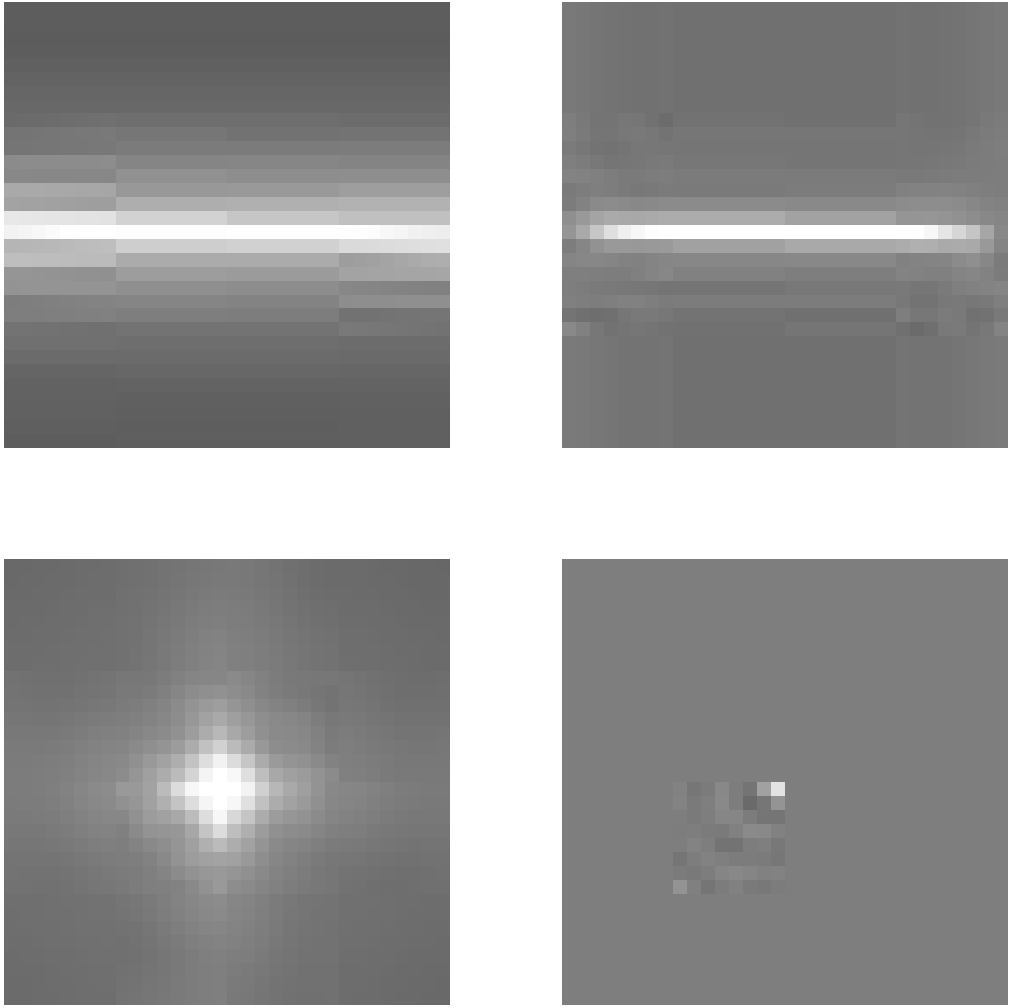}}}}
\caption{
The correlation between the various pixels and the one in the center is 
plotted for the serpentine (upper left), grating (upper right), 
fence (lower left) and random (lower right) scan patterns for the case 
of pure $1/f$ noise.
}
\label{CorrMapsFig}
\end{figure}

The random scan pattern is seen to produce a beautiful white noise power
spectrum, indistinguishable in shape from the above-mentioned
white noise spectrum plotted beneath it. This is quantitative verification
of the claim \cite{Wright96} that a well-connected ``messy" scan differencing
widely separate parts of the sky produces a map with virtually 
uncorrelated pixel noise. This is also seen in \fig{CorrMapsFig}, which shows
how correlated different map pixels are with the one in the center.
Numerical inspection of the covariance matrix $\NN$ shows that it is to a good
approximation proportional to the identity matrix, with the mean of all rows and columns 
subtracted off due to the monopole removal.
Both the fence and grating scans have roughly
\beq{OneOverEllEq}
\Cnoise\propto \l^{-1}
\eeq
over the range of scales probed by the experiment. 
In other words, their angular power spectrum obeys the same
power law in $\l$ as their time power spectrum does in $f$.
Thus although the {\rms} noise per pixel (which 
is dominated by the contribution from $\l$ around 
the pixel separation scale, where the three power spectra
are comparable in magnitude) 
are quite similar for the grating, fence and random scans,
the first two give substantially more power than the third on
larger scales.
This is because there are no ``short cuts'' from one part of
the map to the other, so that large-scale drifts inevitably
leak from the time stream into the spatial noise distribution.
Another way of interpreting this excess large-scale noise is that
although the pixel {\rms} may be small, neighboring pixels are 
correlated so that the effective number of independent pixels is reduced.

Small-scale connectedness helpful as well, as the figure
shows. The four power spectra rank in the same order 
as their degree of connectedness on all scales (the only exception
being the grating scan, where the small-scale noise is raised since half
of the time was spent on the side bars). The serpentine scan is a 
particularly poor performer, with a full order of magnitude 
more noise power than the fence scan on most scales.
The source of the problem with the serpentine scan is illustrated in
\fig{CorrMapsFig}. The fence scan would produce correlation
stripes shaped like a $+$ symbol if the map were
made by simply averaging the observations of each pixel, as is optimal
for white noise. Because of the high degree of interconnectedness, however,
the data analysis method is able to eliminate this striping,
and the correlation region is seen to be fairly round. For the 
serpentine scan, however, the correlation stripe along the scan path
persists. This is because, as is easy to show, the matrix $\W$ 
of \eq{WdefEq} becomes 
the same as for the white noise case in the absence of interconnections, 
apart from removing an overall drift over the entire serpentine. 
In other words, 
the sophisticated mapmaking method is powerless against $1/f$ noise
when the scan pattern is poorly connected.

The situation is seen to be rather intermediate for the
grating scan: the correlation is strong along the scan 
path until it reaches the side bars, where it gets
connected with all the other rows.

\subsection{White and $1/f$ noise combined}

\Fig{NoisePowerFig2} shows the noise power spectra resulting
from a combination of white and $1/f$ noise, where the knee
frequency is a tenth of the sampling rate. Once again, the 
better connected scan strategies are seen to produce less noise power,
although the fence scan is actually very marginally better at the smallest
scales.
The random scan is seen to produce white noise as usual, whereas
the logarithmic slope of $\Cnoise_\l$ for the other scan strategies is
intermediate between the $1/f$ and white cases of $-1$ and $0$.
(We have omitted the grating case to avoid over-crowding this plot.)
The noise power from maps containing 
the white and $1/f$ components alone are also plotted here for comparison,
and it should be noted that the total power when both are present
is always slightly greater than the sum of these curves 
(even though all noise components of course add when $\W$ is held 
fixed), since we cannot optimize for two different types of noise
at the same time.

\subsection{Lesson 3: how to choose the scan strategy}

Connectedness is clearly desirable since it reduces the contribution
from $1/f$ noise to $\Cnoise_\l$. It is also useful for reducing the
susceptibility to systematic errors \cite{Wright96},
and makes the  maps easier to analyze by making
noise correlations more isotropic. However,
complicated interconnected scan patterns can also create 
problems. They might complicate the experimental design, perhaps 
requiring additional moving parts which can cause systematic problems.
For ground and balloon based experiments, a strategy requiring
scans with non-constant elevation can introduce systematic modulations,
since the amount of atmosphere that the beam must penetrate will vary with
time. The relevant question is therefore how great efforts it is
worth expending to increase the connectivity of the scan pattern.

\begin{figure}[phbt]
\centerline{{\vbox{\epsfxsize=16cm\epsfbox{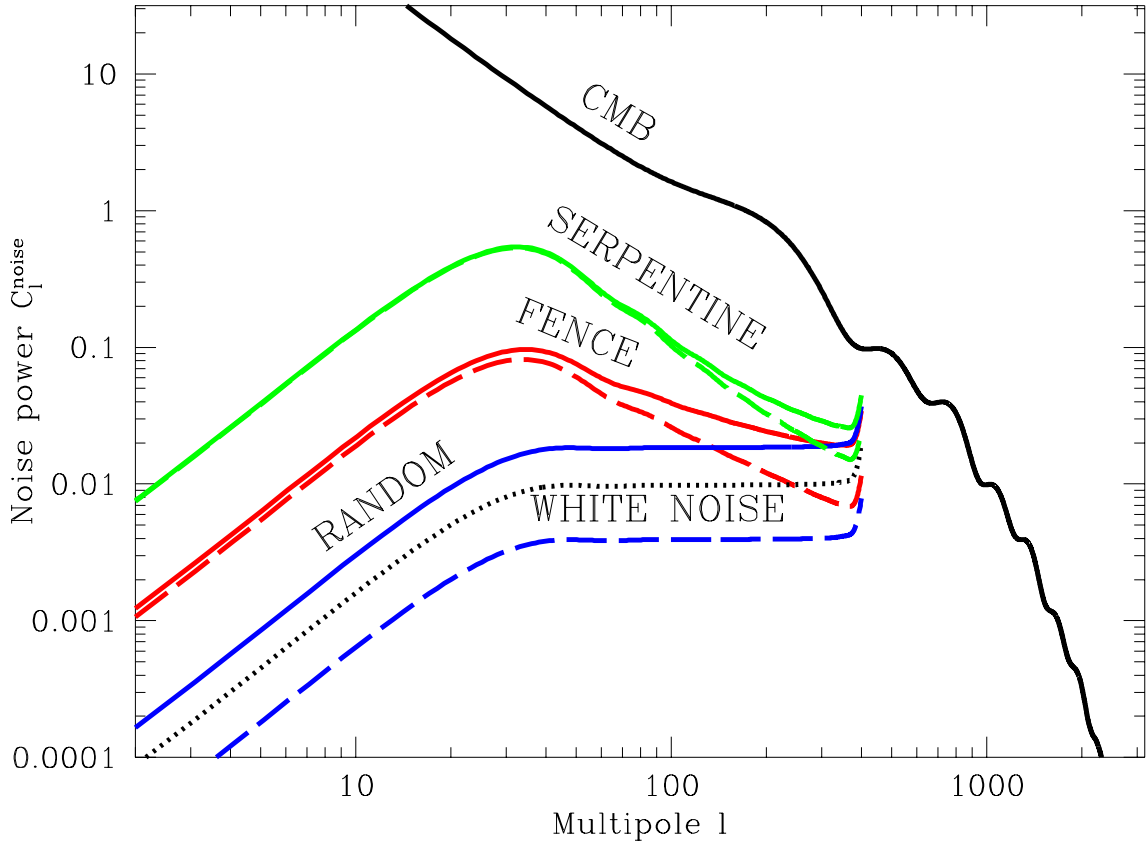}}}}
\caption{
The noise power spectrum $\Cnoise_\l$ is plotted for the
serpentine, fence and random scans with a combination of 
white and $1/f$ noise (solid curves), only the $1/f$ component (dashed
curves) and only the white component (dotted curve, identical for all
three scan patterns).
A standard CDM power spectrum is plotted for comparison (top). }
\label{NoisePowerFig2}
\end{figure}

\Eq{FisherEq} shows that to accurately constrain cosmological
models, we want to minimize the variance $(\Delta\C_\l)^2$ for each multipole.
As we discussed in Section~\ref{SizeSec}, it is
best to chose the map size so that noise and sample variance contribute
roughly equally to $\Delta\C_\l$ on
the pixel scale, {\ie}, at the right edge of the curves in 
\fig{NoisePowerFig2}.
The sample variance scales with $\l$ like the CMB power spectrum,
which is included in \fig{NoisePowerFig2} for comparison.
It lacks the familiar rise at the Doppler peaks simply because we
have plotted $C_\l$ rather than the more familiar quantity 
$\l(\l+1)C_\l$. 
Since we found that $\Cnoise_\l$ never falls
of faster than $\l^{-1}$ (this was for pure $1/f$ noise, and 
a white noise component further reduces the slope) whereas 
$C_\l\simpropto \l^{-2}$, $\Delta C_\l$ will be almost 
completely dominated by sample variance for all but the largest
$\l$-values probed. This means that the huge visual
differences between the noise power spectra are in
fact relatively unimportant
when it comes to measuring cosmological parameters,
with the only really important quantity being the power
on the pixel scale, which is roughly proportional
to the pixel variance $\sigma^2$. $\sigma$ turns out to to be
only 16\% smaller for the random scan than for the fence scan
for pure $1/f$ noise,
and when we included white noise, the fence scan was 
actually the marginally better one (by 6\%).
Our only clearly undesirable scan is the 
serpentine option, which adds noise power 
even on the smallest scales and 
whose {\rms} pixel noise is almost a factor
of two worse than the fence and random scans (this ratio
of course depends strongly on the knee frequency $f_k$).

In conclusion, it is desirable to invest a moderate but not extreme
effort into making the scan pattern more connected than the technically
most convenient option. For a ground- or balloon-based  
experiment, a serpentine-like scan can be readily made
more fence-like by moving the sky patch to be mapped 
further away from the equator, so that 
repeated scans at constant elevation will cross due to 
Earth's rotation. Likewise, a grating-like {\it Planck} great circle 
scan pattern can be made more fence-like by reducing the
angle between the beam and spin axis,
and still more by occasionally tilting the spin axis out of the 
ecliptic plane.  
On the other hand,
going beyond fence connectivity, where one already has nice isotropic
pixel noise with good systematics cross-checks, does probably not
warrant the effort unless it can be done in a technically elegant way such
as for {\it MAP} that does not introduce new potential 
systematic problems.

\section{CONCLUSIONS}

We have discussed the various tradeoffs faced when designing a CMB 
mapping experiment from the point of view of maximizing the scientific 
``bang for the buck''.

Although the traditional approach to this problem has been numerically
expensive Monte Carlo  simulations, we have taken a no-simulation
approach. We found that although state-of-the art data analysis
techniques such as signal-to-noise eigenmode analysis and 
minimum-variance power spectrum estimation are normally treated
as black-box methods, their results can often be accurately approximated
by simple analytic expressions. This allows an intuitive 
understanding of how changing the various experimental parameters
affects the ability to constrain cosmological models.
Illustrating these causal relationships with simple 
case studies, we arrived at the following rules of thumb.
\begin{itemize}
\item {\bf Size:} For a given resolution and sensitivity,
it is best to cover a sky area such that the signal-to-noise ratio
per resolution element (pixel) is of order unity.
\item {\bf Shape:} It is best to avoid excessively ``skinny"
observing regions, narrower than a few degrees.
\item {\bf 1/f-noise:} Scan strategies of both the fence type and the
random type allow the map-making algorithm to substantially
reduce the effect of $1/f$ noise, which makes the noise correlations more isotropic
and produces
a noise power spectrum of slope between $\l^0$ and $\l^{-1}$.
Since this is much flatter than the true CMB spectrum is expected to be, 
slight large-scale noise modulations are cosmologically unimportant
when the map size is chosen as suggested above,
being dwarfed by sample variance.
\end{itemize}

\bigskip
The author wishes to thank Steve Meyer, Phil Lubin
and John Staren for asking questions
that stimulated this work, Uro\v s Seljak and Matias Zaldarriaga for use
of their CMBFAST Boltzmann code, and Ned Wright, the referee, for helpful 
comments.
Support was provided by
NASA through a Hubble Fellowship,
{\#}HF-01084.01-96A, awarded by the Space Telescope Science
Institute, which is operated by AURA, Inc. under NASA
contract NAS5-26555.

\section*{Appendix A: The noise power spectrum with incomplete sky coverage}

In this Appendix, we derive \eq{ClNoiseEq}. Knox
first treated the case $\fsky=1$ \cite{Knox95,wiener},
and early incorrect generalizations to the general case were
corrected by Magueijo \& Hobson \cite{MH97}.
Since no detailed derivation has yet been published for 
this case, we provide one here for completeness.

When the pixel noise is uniform and uncorrelated, 
the quantities $n_i$ (the noise in the  $i^{th}$ pixel)
are random variables satisfying
\beq{NoiseCorrEq1}
\expec{n_i n_j} = \delta_{ij}\sigma^2.
\eeq
Ignoring beam smoothing for the moment, we want to show that 
the effect of this discrete pixel noise is the same as if there 
were a continuous
random field $n(\rh)$ on the sky with power spectrum 
$\Cnoise_\l=C\equiv\Omega\sigma^2/N$. This is a white noise 
power spectrum, since it is independent of $\l$, 
which corresponds to a Dirac delta correlation function
\beq{NoiseCorrEq2}
\expec{n(\rh)n(\rh')} = C\delta(\rh,\rh').
\eeq
As long as the pixelization is uniform 
and fine enough, we can approximate the integral of any function $f$ 
over our patch (of solid angle $\Omega$) by a sum: 
\beq{IntegralApproxEq}
\int f(\rh)d\Omega\approx {\Omega\over N}\sum_{i=1}^N f(\rh_i).
\eeq
When performing a statistical analysis of a CMB map
(for instance a signal-to-noise eigenmode analysis), we
always expand it in some functions, say $\psi,\psi',...$,
so the noise in these expansion coefficients, say $a,a',...$,
is given by 
\beq{zDefEq}
a={\Omega\over N}\sum_{i=1}^N \psi(\rh_i)n_i.
\eeq
For Gaussian noise, the statistical properties of these coefficients
are completely specified by their covariance, which 
using \eq{NoiseCorrEq1} is given by
\beqa{aCovEq1}
\expec{aa'} 
&=& \left({\Omega\over N}\right)^2
\sum_{i=1}^N \sum_{j=1}^N \psi(\rh_i)\psi'(\rh_j)\expec{n_i n_j}\nonumber\\
&=& \left({\Omega\sigma\over N}\right)^2
\sum_{i=1}^N \psi(\rh_i)\psi'(\rh_i)
\eeqa
When computing this covariance as if the noise where a 
continuous random field, \eq{IntegralApproxEq} gives
\beq{zDefEq2}
a\approx\int\psi(\rh)n(\rh)d\Omega,
\eeq
and using \eq{NoiseCorrEq2}, we obtain
\beqa{aCovEq2}
\expec{aa'} 
&\approx& \int\int\psi(\rh)\psi'(\rh')\expec{n(\rh)n(\rh')}d\Omega d\Omega'\nonumber\\
&=& C\int \psi(\rh)\psi'(\rh) d\Omega\\
&\approx& {\Omega C\over N}\sum_{i=1}^N \psi(\rh_i)\psi'(\rh_i).
\eeqa
Comparing equations\eqno{aCovEq1} and\eqno{aCovEq2}, we see
that the two ways of treating the noise give 
the same answer if $C=\Omega\sigma^2/N$.
All that remains to prove \eq{ClNoiseEq} is to divide the right hand side by 
the beam correction $B_\l^2$, noting that the noise is added to the sky signal 
{\it after} it has been smoothed by the experimental beam.




\onecolumn
\clearpage

\clearpage

\clearpage

\label{RectModesFig}
\clearpage

\clearpage

\clearpage

\clearpage

\clearpage

\clearpage

\clearpage

\clearpage

\clearpage

\clearpage

\end{document}